\RequirePackage{fix-cm}
\documentclass[smallextended]{svjour3}       
\smartqed  
\usepackage{graphicx}

\usepackage{natbib}
\usepackage{listings}
\usepackage{xspace}
\usepackage{multirow}
\usepackage[table,xcdraw]{xcolor}
\usepackage{enumitem}
\usepackage[caption=false,font=footnotesize]{subfig}
\usepackage[switch]{lineno}

\usepackage{float}
\usepackage{rotating}
\usepackage{rotating}

\usepackage{caption}

\usepackage{booktabs} 
\usepackage{siunitx} 
\usepackage{tabularx} 
\usepackage{adjustbox}
\usepackage{multicol}

\definecolor{codegreen}{rgb}{0,0.6,0}
\definecolor{codegray}{rgb}{0.5,0.5,0.5}
\definecolor{codepurple}{rgb}{0.58,0,0.82}
\definecolor{backcolour}{rgb}{0.95,0.95,0.92}

\lstdefinestyle{mystyle}{
	basicstyle=\ttfamily\fontsize{8}{8}\selectfont,
	breakatwhitespace=false,         
	breaklines=true,                 
	captionpos=b,                    
	keepspaces=true,                 
	numbersep=5pt,                  
	showspaces=false,                
	showstringspaces=false,
	showtabs=false,                  
	tabsize=4
}

\lstdefinestyle{pstyle}{
	commentstyle=\color{codegreen},
	keywordstyle=\color{magenta},
	numberstyle=\tiny\color{codegray},
	stringstyle=\color{red},
	basicstyle=\ttfamily\fontsize{8}{8}\selectfont,
	breakatwhitespace=false,         
	breaklines=true,                 
	captionpos=b,                    
	keepspaces=true,                 
	numbers=left,                    
	numbersep=5pt,                  
	showspaces=false,                
	showstringspaces=false,
	showtabs=false,                  
	tabsize=4
}

\usepackage{tikz}
\usetikzlibrary{shapes, shapes.geometric, arrows, positioning}

\usepackage{flushend}
\usepackage{dirtree,array}
\usepackage{fancybox, graphicx}

\usepackage[colorinlistoftodos,prependcaption,textsize=tiny]{todonotes}

\usepackage[hidelinks=true
pdftex,
pdfauthor={Ashwin Prasad Shivarpatna Venkatesh, Samkutty Sabu, Mouli Chekkapalli, Jiawei Wang, Li Li, Eric Bodden},
pdftitle={Static Analysis Driven Enhancements for Comprehension in Machine Learning Notebooks},
pdfkeywords={static analysis, python, code comprehension, annotation, literate programming, jupyter notebook}
]{hyperref}

\begin{document}

\title{Static Analysis Driven Enhancements for Comprehension in Machine Learning Notebooks\thanks{Funding for this study was provided by the Ministry of Culture and Science of the State of North Rhine-Westphalia under the SAIL project with the grant no NW21-059D.}
}

\titlerunning{Static Analysis Driven Enhancements for Comprehension in ML Notebooks}        

\author{Ashwin Prasad Shivarpatna Venkatesh \and
Samkutty Sabu \and
Mouli Chekkapalli \and
Jiawei Wang \and
Li Li \and
Eric Bodden
}

\institute{
Ashwin Prasad Shivarpatna Venkatesh\and Samkutty Sabu \and Mouli Chekkapalli\at
  Heinz Nixdorf Institut, Paderborn University, Paderborn, Germany\\                         
  \email{ashwin.prasad@upb.de}\\
  \email{samkutty@mail.uni-paderborn.de, moulik@mail.uni-paderborn.de}
\and
Jiawei Wang \at
Faculty of Information Technology, Monash University, Melbourne, Australia\\
\email{jiawei.wang1@monash.edu}
\and
Li Li \at
School of Software, Beihang University, Beijing, China\\                    
\email{lilicoding@ieee.org}
\and
Eric Bodden \at
Heinz Nixdorf Institut \& Fraunhofer IEM, Paderborn University, Paderborn, Germany\\                         
\email{eric.bodden@upb.de}
}

\date{Received: date / Accepted: date}

\newcommand{\python}{Python\xspace}
\newcommand{\jnb}{notebook\xspace}
\newcommand{\code}[1]{\texttt{#1}}
\newcommand{\toolname}{\textsc{HeaderGen}\xspace}
\newcommand{\typeevalpyrawname}{TypeEvalPy\xspace}
\newcommand{\typeevalpy}{\textsc{TypeEvalPy}\xspace}

\maketitle

\begin{abstract}
Jupyter notebooks have emerged as the predominant tool for data scientists to develop and share machine learning solutions, primarily using Python as the programming language.	
Despite their widespread adoption, a significant fraction of these notebooks, when shared on public repositories, suffer from insufficient documentation and a lack of coherent narrative.
Such shortcomings compromise the readability and understandability of the notebook.
Addressing this shortcoming, this paper introduces \toolname, a tool-based approach that automatically augments code cells in these notebooks with descriptive markdown headers, derived from a predefined taxonomy of machine learning operations. Additionally, it systematically classifies and displays function calls in line with this taxonomy.

The mechanism that powers \toolname is an enhanced call graph analysis technique, building upon the foundational analysis available in \emph{PyCG}. 
To improve precision, \toolname extends PyCG's analysis with return-type resolution of external function calls, type inference, and flow-sensitivity.
Furthermore, leveraging type information, \toolname employs pattern matching techniques on the code syntax to annotate code cells.

We conducted an empirical evaluation on 15 real-world Jupyter notebooks sourced from Kaggle. The results indicate a high accuracy in call graph analysis, with precision at 95.6\% and recall at 95.3\%. 
The header generation has a precision of 85.7\% and a recall rate of 92.8\% with regard to headers created manually by experts.
A user study corroborated the practical utility of \toolname, revealing that users found \toolname useful in tasks related to comprehension and navigation.

To further evaluate the type inference capability of static analysis tools, we introduce \typeevalpy, a framework for evaluating type inference tools for Python with an in-built micro-benchmark containing 154 code snippets and 845 type annotations in the ground truth.
Our comparative analysis on four tools revealed that \toolname outperforms other tools in exact matches with the ground truth. 

\end{abstract}

\section{Introduction}
\label{sec:intro}
In the evolving landscape of  machine learning (ML) and data science, Jupyter Notebooks have emerged as the predominant platform for creating ML solutions within the ML community.
These notebooks resonate with the paradigm of \textit{literate programming} postulated by \cite{knuthLiterateProgramming1984a}. This approach advocates the integration of code, comprehensive documentation, and visual representations within a unified document to foster understanding and facilitate the sharing  of intricate solutions.
The underlying principles of literate programming include:
(1) Augmenting code with descriptive text and illustrative diagrams.
(2) Imposing a coherent narrative by separating code cells with pertinent headers.
(3) Logically segmenting and labeling the program's reusable modules.

Within notebooks, code is written in executable \textit{code cells} while accompanying documentation is written in \textit{markdown cells}.
An illustrative example of this configuration, showcasing \python{} code cells interleaved with markdown cells, is depicted in Figure \ref{motivatingnb}. 
Note that \python{} has emerged as the predominant language of choice for articulating ML-based solutions in notebooks~\citep{ruleExplorationExplanationComputational2018}.

Augmenting code segments with explanatory textual content enhances the overall comprehensibility of notebooks and further promotes collaboration~\citep{wagemann2022five}.
Moreover, a markdown-to-code cell ratio of 2, as posited by \citet{wagemann2022five}, serves as an indication of commendable literate programming adherence. 
This assertion finds further support in the work of \cite{samuel2022computational}, who report that a higher markdown-code cell proportion correlates with superior reproducibility, a vital metric in scientific studies.

However, despite the inherent capabilities of Jupyter \jnb{}s resonant with literate programming principles, real-world adoption often diverges from this ideal~\citep{keryStoryNotebookExploratory2018}. 
Empirical studies reveal that code-smells and suboptimal practices are common in publicly distributed notebooks~\citep{wangBetterCodeBetter2020}. 
Interviews conducted by \cite{ruleExplorationExplanationComputational2018} with ML practitioners showed that they often defined their notebooks as personal unstructured scratch-pads and ``messy''.
The authors found that the reluctance in annotating notebooks has been attributed to time constraints or being ``too lazy''.
Subsequent large-scale research by \cite{pimentelLargeScaleStudyQuality2019} underscored that 30.93\% of 1.4 million notebooks they examined lack markdown cells, a finding corroborated by \cite{quarantaBestPractices2022}. 
This scarcity of annotations reveals the prevailing negligence towards best practices.
Such omissions are especially detrimental in platforms like Kaggle\footnote{\url{https://www.kaggle.com/}}, where subpar practices risk propagating to the next generation of ML practitioners.

Therefore, there exists an imperative for the software engineering research community to develop tools that address this problem.
To this end, this paper introduces \toolname{}, a tool-based approach to augment the comprehension and navigation of undocumented \python{} based Jupyter notebooks by automatically creating a narrative structure in the notebook.

\begin{figure}
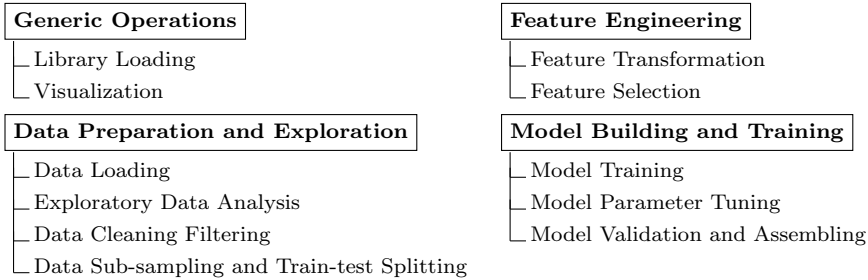

	\begin{minipage}[t]{0.55\textwidth}
\renewcommand\DTstyle{\rmfamily}
\DTsetlength{0.2em}{0.7em}{0.2em}{0.4pt}{0pt}
\dirtree{%
	.1 \fbox{\textbf{Generic Operations}}\vspace{-.6ex}.
	.2 Library Loading.
	.2 Visualization.
}

\vspace{.7em}

\DTsetlength{0.2em}{0.7em}{0.2em}{0.4pt}{0pt}
\dirtree{%
	.1 \fbox{\textbf{Data Preparation and Exploration}}\vspace{-.6ex}.
	.2 Data Loading.
	.2 Exploratory Data Analysis.
	.2 Data Cleaning Filtering.
	.2 Data Sub-sampling and Train-test Splitting.
}

	\end{minipage}%
	\begin{minipage}[t]{0.45\textwidth}
		\renewcommand\DTstyle{\rmfamily}
\DTsetlength{0.2em}{0.7em}{0.2em}{0.4pt}{0pt}
\dirtree{%
	.1 \fbox{\textbf{Feature Engineering}}\vspace{-.6ex}.
	.2 Feature Transformation.
	.2 Feature Selection.
}

\vspace{.7em}

\DTsetlength{0.2em}{0.7em}{0.2em}{0.4pt}{0pt}
\dirtree{%
	.1 \fbox{\textbf{Model Building and Training}}\vspace{-.6ex}.
	.2 Model Training.
	.2 Model Parameter Tuning.
	.2 Model Validation and Assembling.
}

	\end{minipage}
	\caption{Taxonomy of machine learning operations based on \citep{wangDocumentationMattersHumanCentered2022a}.}
	\label{t:ml_operations}
\end{figure}

The work of \cite{wangDocumentationMattersHumanCentered2022a} has shed light on the structured taxonomy of ML operations.
As demonstrated in Figure~\ref{t:ml_operations}, the process undertaken by data scientists when developing an ML-based notebook typically begins with data preparation, followed by feature extraction, and continues with the creation and training of the model.
Within this structured process lies an implicit narrative structure that is key to the operation of \toolname. 
We specifically designed \toolname to capture this narrative by accurately detecting each function call in the ML notebook.
Subsequently, it classifies each call according to the ML operations taxonomy.
Utilizing this classified data, \toolname creates a structural map of the entire notebook which is then presented as an \textit{``index of ML operations''} at the top of the notebook.
This is further complemented by annotating every code cell with relevant markdown headers that highlight the ML operations being performed (see example in page \pageref{fig:index_ml_ops}).

To yield useful results, \toolname{} requires a fast and accurate program analysis that can precisely identify all function calls in the notebook.
An exploration into existing methodologies revealed an absence of a technique that could statically identify all function calls in a \jnb{} with acceptable precision, recall, and execution time.
This can be attributed to Python's inherent complexities such as duck typing, dynamic code execution, reflection, among others that are challenging to static analysis~\citep{salisPyCGPracticalCall2021a,kummitaQualitativeQuantitativeAnalysis2021}.
Moreover, in contrast to other programming languages like Java, \python{} lacks a lot of tool-support for state-of-the-art static analysis (SA) techniques \citep{yangComplexPythonFeatures2022}.
The prevailing tools for Python predominantly rely on rudimentary analyses of its abstract syntax trees (AST).
Compounding this issue is Python's dynamically-typed nature, necessitating a precise static type-inference for variables to facilitate accurate static analysis.
A recent advancement in the form of a call-graph generation technique, PyCG~\citep{salisPyCGPracticalCall2021a}, which operates on an intermediate representation of the AST and addresses many complex Python features, also falls short.
Specifically, PyCG ignores analyzing function calls to external libraries and presents a flow-\textit{insensitive} analysis.
Flow sensitivity in SA refers to the ability of the analyzer to consider the order in which statements occur in the program.
A flow-sensitive analysis tracks the flow of control through the program.
This means the analysis can differentiate between the different states of variables at different points in the program.
Flow-insensitivity of PyCG limits its capability to accurately extract function calls in real-world notebooks.
\toolname rectifies these limitations.

To summarize, the challenge that \toolname addresses is two-fold:
\begin{enumerate}
	\item \textbf{Inadequacies in Static Analysis:} the absence of a static program analysis technique that can precisely identify function calls in a \python program.
	To mitigate this, \toolname extends the analysis in PyCG with the ability to resolve function calls to external libraries, infer types of variables, and adds flow-sensitivity.
	\item \textbf{Lack of Documentation in Notebooks:} a considerable portion of publicly accessible \jnb{}s lack adequate documentation. This absence not only impedes comprehension but also violates the principles of literate programming. \toolname employs precise static analysis to automatically augment these notebooks with structural headers. This addition creates a narrative structure to aid comprehension of undocumented \jnb{}s.
\end{enumerate}

To evaluate the performance of \toolname's static function call analyzer, we employed an enhanced version of PyCG's micro-benchmark, complemented by a real-world benchmark consisting of 15 \jnb{}s sourced from Kaggle.
On the real-world benchmark, \toolname achieved 95.6\% precision and 95.3\% recall, outperforming PyCG and other function call analyzers based on off-the-shelf tools such as \emph{pyright}\footnote{\url{https://github.com/microsoft/pyright}} and \emph{Jedi}\footnote{\url{https://github.com/davidhalter/jedi}}.
On the same benchmark we also evaluated \toolname for header annotation and achieved 85.7\% precision and 92.8\% recall.
Additionally, through a user study involving eight data science practitioners, we gathered evidence indicating that \toolname significantly enhances navigation speed and improves comprehension.

Static function call analysis is dependent on the analyzer's capability to statically infer types of variables.
Therefore, we systematically evaluated the type inference capabilities of these tools.
To facilitate this, we created, \typeevalpy, the first micro-benchmarking framework specifically designed for type inference evaluation in Python.
\typeevalpy contains 154 code snippets, organized into 18 categories, each focusing on distinct language-specific features of Python, summing up to a comprehensive 845 type annotations.
Notably, \toolname registered 564 exact matches against the 845 annotations in the ground truth, a performance that notably surpasses other tools.

The primary contributions of our work are as follows:

\begin{itemize}

\item We propose a novel static analysis based approach for \python Jupyter notebooks that can automatically enhance them with structural, explanatory, and navigational annotations to augment literal programming practice.
\item We implement a static function call extraction technique for \python{} with 95.6\% precision and 95.3\% recall on our real-world benchmark.
\item We give an evaluation of our approach based on extensive experimental results.
\item We implement the prototype named \toolname and make it publicly available for our community to reuse. 
\end{itemize}

This manuscript builds upon the findings presented in our initial publication \citep{venkateshEnhancingComprehensionNavigation2023c}.
Here, we detail the incremental advancements made since our initial publication:

\begin{itemize}
\item We introduce \typeevalpy, the first micro-benchmarking framework for type inference evaluation in Python containing 154 code snippets and 845 manually annotated types.

\item A detailed evaluation of several type inference tools using \typeevalpy, including \toolname. This assessment demonstrates how improved type inference enhances callsite recognition and header generation in Jupyter notebooks.

\item The original dataset is expanded to include 15 new Jupyter notebooks from real-world projects, involving detailed manual annotations to identify headers and function callsites, and to extract fully qualified names. This validates \toolname's performance on a broader and more diverse dataset.

\item To overcome the limitation in the original \toolname approach concerning the manual curation of a database that maps ML API calls to an ML taxonomy, we developed an ML model that classifies function calls using their documentation strings. Trained on a dataset curated by four data science experts, this model enhances \toolname's adaptability to evolving ML libraries and functions.

\end{itemize}

The remainder of this paper is organized as follows: we present challenges in statically analyzing Python code with a motivating example in the Section~\ref{sec:motivating_example}, followed by detailing our design in Section~\ref{sec:method}. 
We then present the research questions in Section~\ref{sec:evaluation}.
We discuss the details of our micro-benchmark, real-world benchmark, and \typeevalpy in Section~\ref{sec:benchmarks}.
We address the research questions from Section~\ref{sec:userstudy} to Section~\ref{subsec:RQ5}, and discuss existing research in Section~\ref{sec:relatedwork}.
The limitations of \toolname is discussed in Section~\ref{sec:discussion} and finally the paper is concluded in Section~\ref{sec:conclusion}.

\textbf{Availability.}
\toolname is published on GitHub as open-source software under Apache 2.0 license:\\ \url{https://github.com/secure-software-engineering/HeaderGen}.

\typeevalpy is published on GitHub as open-source software under Apache 2.0 license: \url{https://github.com/secure-software-engineering/TypeEvalPy}.

\section{Motivating Example}
\label{sec:motivating_example}

\begin{figure}[t]
	\centerline{\includegraphics[width=.95\linewidth]{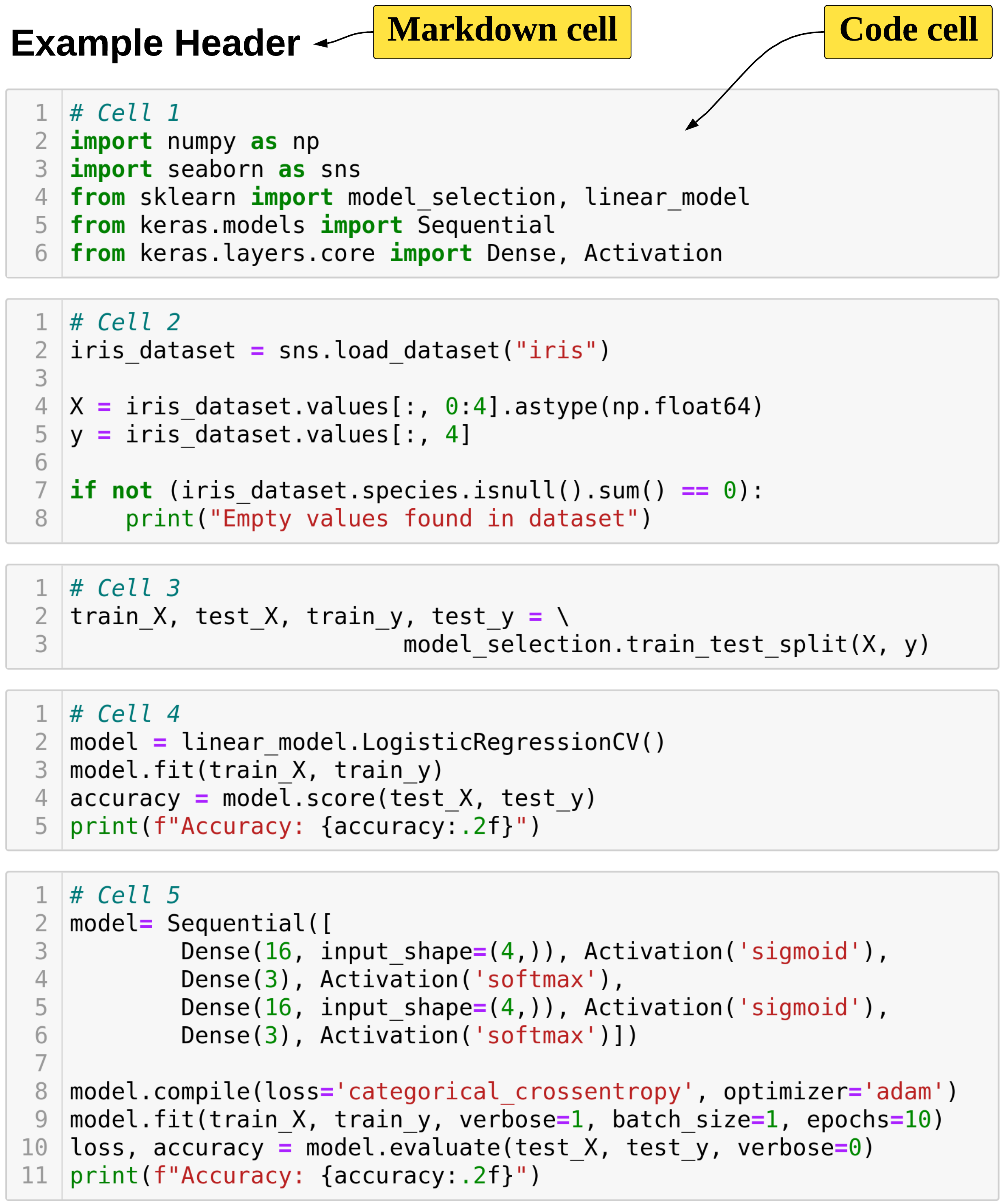}}
	\caption{Example of a Jupyter notebook containing a machine learning  solution.}
	\label{motivatingnb}
\end{figure}

As a motivating example, consider the \jnb{} in Figure~\ref{motivatingnb}.
It consists of one markdown cell which is rendered as an HTML header, and five code cells that can be identified by the comments in the first line of each code cell.
The example notebook in Figure \ref{motivatingnb} is a concise version of a real-world notebook containing a machine learning (ML) based solution.

In cell 1, various ML libraries are imported.
In cell 2, a sample dataset called ``iris" from the seaborn library is loaded, and further feature selection operations are performed to retain only the essential columns from the dataset.
Values are type-cast to numpy based \texttt{float64} type.
Finally, the dataset is checked for null values.
In cell 3, the dataset is split into training and test datasets.
In cells 4 and 5, with the processed dataset, two different ML models are defined, trained, and their accuracies are reported. 
In cell 4, a basic linear model based on logistic regression is used.
In cell 5, a deep learning based sequential model is used.

Note that this notebook is undocumented and does not contain any explanatory text or structural headers as markdown cells, violating the literate programming principle.
One in three notebooks found in the wild does not contain any markdown cells~\citep{pimentelLargeScaleStudyQuality2019}.
In absence of explanatory text or structural headers, ML~practitioners, especially beginners, must spend more time to navigate and comprehend different aspects of the notebook.
Particularly considering that nearly a third of all notebooks in the real-world contain at least 50 cells~\citep{pimentelLargeScaleStudyQuality2019}.

On the other hand, the example \jnb{} poses several challenges to SA, including:

\begin{itemize}[leftmargin=0.2cm]
	\item \textbf{Import aliasing:} different ways of importing libraries, and importing libraries with aliases. This is further complicated with wild card imports of the form ``\texttt{from MODULE import *}''.
	
	\item \textbf{Dynamic typing:} in cell 2, the type of the variable \texttt{iris\_dataset} is not known statically, i.e., the return-type of the function \texttt{load\_dataset()} is not known statically unless the developer manually annotates the function definition.
 	Unfortunately, widespread adoption of type annotation is still lacking in practice \citep{digraziaEvolutionTypeAnnotations2022a}.
	As a result, subsequent statements that involve the variable \texttt{iris\_dataset} cannot be resolved, i.e., in cell~2 lines~4--7.
	
	\item \textbf{Chained function calls:} consider the function call in cell~2 line~4, \\ \texttt{iris\_dataset.values[].astype()}, here, the variable \texttt{iris\_dataset} is of type \textit{Dataframe} from the \textit{Pandas} library. \texttt{iris\_dataset.values} refers to an attribute of the class \textit{Dataframe}, which is in-turn defined as a \textit{Numpy} array.
	Furthermore, \texttt{astype()} refers to a function from the \textit{Numpy} library.
	While the syntactic chaining requires careful modeling by the analyzer, the main challenge is that the imprecision of analysis is propagated through these call chains.
	Existing SA tools fail to resolve all this information statically.
	
	\item \textbf{Variable reuse:} the same variable \texttt{model} is reused in cells 4 and 5, for different model objects, i.e., \texttt{Sequential} and \texttt{LogisticRegressionCV} objects. 
	Reuses of the same variable names are common in \jnb{}s.
	Therefore, for precise annotation of code cells, the analyzer should know the type of an object at a specific location in the notebook.
	In other words, the analysis should be flow-\textit{sensitive}.

\end{itemize}

Note that in general, analyzing dynamic programming languages such as Python poses several other challenges not discussed in our example.
Features such as dynamic evaluation where a string can be evaluated as a code fragment at runtime and complex control-flow structures with generators also pose challenges to SA.

In summary, for \toolname to accurately classify code cells based on function calls, the static analyzer needs to: 
(1)~handle complex \python features,
(2)~statically resolve return-types of external library calls, and
(3)~be flow-sensitive.

\section{Approach}

\label{sec:method}
\begin{figure}[]
	\centerline{\includegraphics[width=1\linewidth]{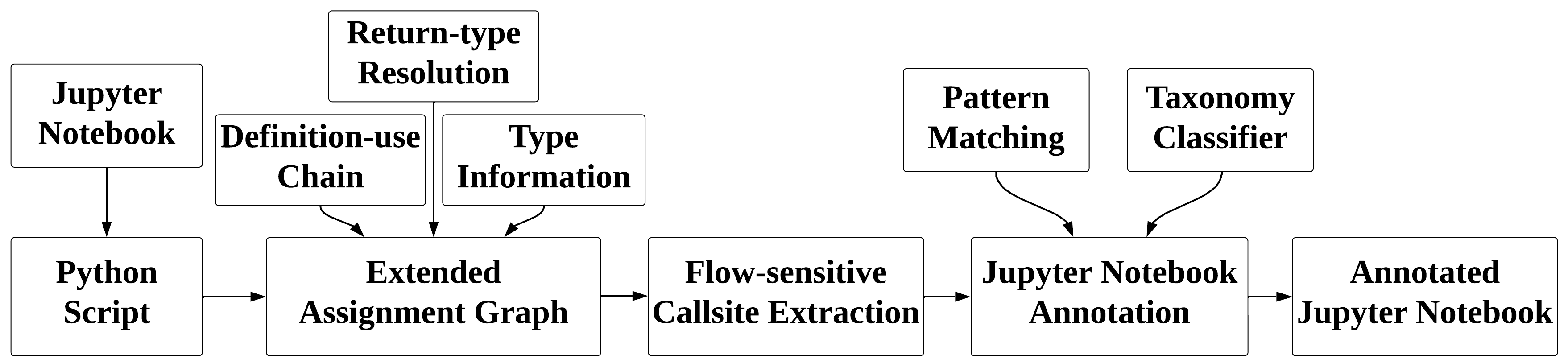}}
	\caption{High-level overview of \toolname{}.}
	\label{fig:overview}
\end{figure}

A comprehensive overview of \toolname{} is shown in Figure \ref{fig:overview}.
The initial step involves transforming a \jnb{} into a standard \python{} script.
This process aids in eliminating metadata from the \jnb, ensuring that only pertinent information remains for subsequent analysis. Post this conversion, \toolname proceeds to analyze the resultant \python{} script to create an extended assignment graph (EAG). Leveraging the EAG, \toolname extracts flow-sensitive callsite information.
The final phase involves enriching the \jnb{} with relevant headers corresponding to the callsites by using an ML-based taxonomy classifier.
Additionally, an index of ML operations is generated.

The details of the EAG construction and the methodology for extracting flow-sensitive callsite information are discussed in Sections \ref{subs:eag} and \ref{subs:flowsens}.
The~ML-based taxonomy classifier is elaborated in Section \ref{sec:taxonomy_classifier}.
Subsequently, Section \ref{subs:jnannotation} outlines the procedure employed for annotating the notebook, utilizing the outputs from the analyzer.

\subsection{Extended Assignment Graph}
\label{subs:eag}
\begin{figure}[]
	\centering
	\subfloat[Assignment Graph]{
		\begin{minipage}[b][0.15\textheight][c]{0.27\linewidth}
			\centering
			\vspace*{\fill}
			\begin{tikzpicture}[scale=1.15, transform shape]
				
				\tikzstyle{oval} = [ellipse, draw, align=center]
				
				\node[oval,anchor=center] (model1) {model};
				
			\end{tikzpicture}
			\vspace*{\fill}
		\end{minipage}
		
		\label{fig:agpycg}
		
	}
	\hfill
	\raisebox{13mm}{\Large $\Rightarrow$}
	\hfill
	\subfloat[Extended Assignment Graph]{
		\begin{tikzpicture}[scale=0.85, transform shape]
			
			\tikzstyle{oval} = [ellipse, draw, align=center]
			
			\node[oval] (model1) {model:\color{red}{(C4,2)}};
			\node[oval, below=.5cm of model1] (sklearn) {sklearn.linear\_model.\\ \_logistic.\\ LogisticRegressionCV};
			\node[oval, right=2cm of model1] (model2) {model:\color{red}{(C5,2)}};
			\node[oval, below=.5cm of model2] (keras) {keras.engine.\\ sequential.Sequential};
			
			\draw[->] (model1) -- (sklearn);
			\draw[->] (model2) -- (keras);
			
		\end{tikzpicture}
		\label{fig:ex_ag}	
	}
	\caption{Generated assignment graphs for the variable ``model'' in the motivating example, left in PyCG (empty), right in HeaderGen (flow-sensitive).}
	\label{fig:af_compare}
\end{figure}
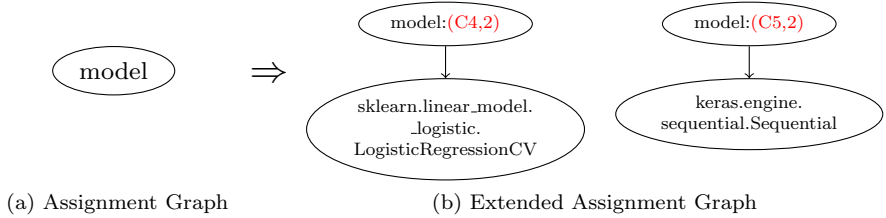

To extract all possible callsites in the program, we add flow-sensitivity and the ability to analyze external libraries to the existing state-of-the-art context-insensitive and inter-procedural call-graph (CG) generation technique, PyCG \citep{salisPyCGPracticalCall2021a}.
PyCG works on a custom intermediate representation of a \python{} AST and generates an assignment graph (AG) that represents assignment relations between program identifiers.
The CG is then generated based on the AG by resolving all function calls that a program variable might point-to.
Figure \ref{fig:agpycg} shows the AG generated by PyCG for the variable \texttt{model} in our motivating example.
Since PyCG cannot analyze calls to external libraries, it does not add any edges to the \texttt{model} node.
However, callsite information from real-world notebooks cannot be extracted with high accuracy without analyzing external library functions.
To wit, in our motivating example, without analyzing the function \texttt{load\_dataset()} from the seaborn library, further references to the variable \texttt{iris\_dataset} cannot be resolved.
Moreover, PyCG's analysis is flow-\emph{insensitive}, therefore the generated AG fails to distinguish between different assignments to the same variable.
For instance, in our motivating example, \texttt{model} is first defined in cell 4 and then redefined in cell 5 (cf. Figure \ref{motivatingnb}), however, the generated AG shown in Figure \ref{fig:agpycg} maintains only a single node for the \texttt{model} variable.
PyCG over-approximates \texttt{model} with weak-updates to the AG, thereby, compromising on precision.

Therefore, we extend PyCG's AG by an extended assignment graph (EAG) based on an additional helper analysis to enable flow-sensitive callsite recognition and further add a return-type approximation technique to resolve calls to external libraries.

\subsubsection{Definition-use Chain for Flow-sensitivity}
A definition-use chain (DUC) \citep{kennedyUsedefinitionChainsApplications1978} is a data structure that represents a definition, or assignment, of a program variable and all the subsequent uses without any re-definitions in between.
DUCs are generated by analyzing all assignment statements in the program with consideration of variable scopes.

We use an existing tool, Beniget \citep{serge-sans-pailleGastBeniget2022}, a DUC generation tool that works by analyzing the AST of \python{} programs.
While a tool exists for \python{} to compute the DUC, no existing implementation makes use of DUC to construct flow-sensitive call-graphs for \python{}.
\toolname first uses the DUC generated by Beniget to create a location map that gives information about what variables are used at particular locations of a notebook.
Then, this map is used to create the EAG that can differentiate variables based on the location of its definition. 
For instance, the EAG shown in Figure \ref{fig:ex_ag} captures multiple definitions of the \texttt{model} variable separately.

\subsubsection{Return-type Resolution of Machine Learning Libraries}
Consider the variable \texttt{iris\_dataset} assigned to the return of function \texttt{load\_dataset()} at location cell 2 line 2, represented as (C2,2) in the motivating example (Figure \ref{motivatingnb}).
Within the seaborn library, the call to \texttt{load\_dataset()} is resolved to \texttt{seaborn.utils.load\_dataset}, which returns an object of type \texttt{pandas.Dataframe}. 
For \toolname, this type information is crucial: only if \toolname knows \texttt{iris\_dataset}'s type can it statically analyze calls on this variable.
For instance \texttt{iris\_dataset} is used at (C2,4), (C2,5), and (C2,7) all of which cannot be resolved without knowing \texttt{iris\_dataset} is of type \texttt{pandas.Dataframe}.
Yet, \python{} is a dynamically typed language, return-type information is not readily available for most library code.
Although a set of Python Enhancement Proposals (PEPs) such as PEP484\footnote{\url{https://peps.python.org/pep-0484/}} are placed for Python language to support type annotations directly in source code, recent work has suggested that such user-demanding knowledge is still missing  \citep{di2022evolution}.

Though it still remains an open challenge, researchers have given type inference for Python a lot of attention.
While leading tech giants like Google, Meta, and Microsoft rely on static tools (e.g. \emph{pytype}~\citep{Pytype2022}) to ensure the quality of their codebase, the majority of current efforts employ the deep learning technique.
Unfortunately, none of the  available tools can accomplish what we need.
This is mainly because external function calls frequently create dataflow disruptions in notebook programs. 
Existing learning-based approaches such as Typilus~\citep{allamanisTypilusNeuralType2020} and Type4Py~\citep{type4py} often only leverage  the source code's contextual information to generate the probabilistic type candidates.
Static tools such as \emph{pytype} and \emph{pyre} often ship with tailored type stubs, providing no support for user type stubs. 
The two tools also do not infer types for local variables, leaving class method calls hard to obtain.
\emph{pyright}~\citep{StaticTypeChecker2022}, a type checking tool, enables support for using custom type stubs of external libraries, but, does not model library specific behavior leading to loss of recall.
Moreover, \emph{pyright} needs to be further re-engineered to obtain inferred type hints as it is designed for type checking \citep{yang2022data}. 
The well-known open-source project \emph{Jedi}~\citep{halterJediAwesomeAutocompletion2022} cannot analyze complex \python features, and suffers from performance issues.
Furthermore, \emph{MOPSA}~\citep{MOPSAMOPSAAnalyzer2024} is a generic static analysis framework with a focus on analyzing Python programs that use external libraries written in C.
It is focused on finding type and value errors.
MOPSA uses formal methods and proposed Python-C semantics to model program states in C and Python side.
However, C Code is rather not dominant (e.g. XGBoost and LightGBM) in our context.
MOPSA, however, focuses on analyzing Python-C code only.
We summarize the usage of cross-language usages in Table \ref{table:mopsa_c_python}.
The percentage of each language usage is retrieved from their official GitHub page.

\begin{table}[]
\renewcommand{\arraystretch}{1.2}
\centering
\caption{Overview of cross-language usage in machine learning libraries}
\label{table:mopsa_c_python}
	\begin{tabular}{lrr}
		\toprule
		\multicolumn{1}{c}{\textbf{Library}} & \multicolumn{1}{c}{\textbf{Python}} & \multicolumn{1}{c}{\textbf{C}} \\ \midrule
		Pandas                                & 90.4                                & 1.6                            \\
		Sklearn                               & 92.2                                & 0.3                            \\
		Matplotlib                            & 92.6                                & 0                              \\
		Numpy                                 & 60                                  & 35.7                           \\
		Seaborn                               & 100                                 & 0                              \\
		Keras                                 & 99.9                                & 0                              \\
		LightGBM                              & 20.4                                & 3.2                            \\
		XGBoost                               & 20.6                                & 0                              \\
		TensorFlow                            & 26.6                                & 0                              \\
		Plotly                                & 98.1                                & 0                              \\
		Torch                                 & 50                                  & 2                              \\
		Statsmodels                           & 92.8                                & 0                             \\
		\bottomrule
	\end{tabular}
\end{table}

\emph{PyCG} is of no help here: it does not analyze calls to external libraries, instead ignores them.
We attempted to force \emph{PyCG} to analyze ML libraries such as \emph{Numpy} and \emph{Pandas}.
Yet, we failed to obtain results due to crashes and out-of-memory exceptions.
External libraries, especially ML libraries, can contain millions of lines of code and PyCG's fixed-point algorithm does not terminate within reasonable time and memory.
Even after (unsoundly) limiting the number of iterations of PyCG's fixed-point algorithm, the resulting AG was unsuitable for real-world application because of the low precision and recall. 
An analysis of the ML libraries' code thus seems out of reach with current tooling.
We further explore these limitations with a quantitative comparison of \emph{PyCG}, \emph{Jedi}, \emph{pyright}, and \emph{HiTyper}~\citep{HiTyper} with \toolname in Sections \ref{sec:eval_typeevalpy} and \ref{subsec:RQ5}.

\begin{figure}[tbp]
	\centerline{\includegraphics[width=.8\linewidth]{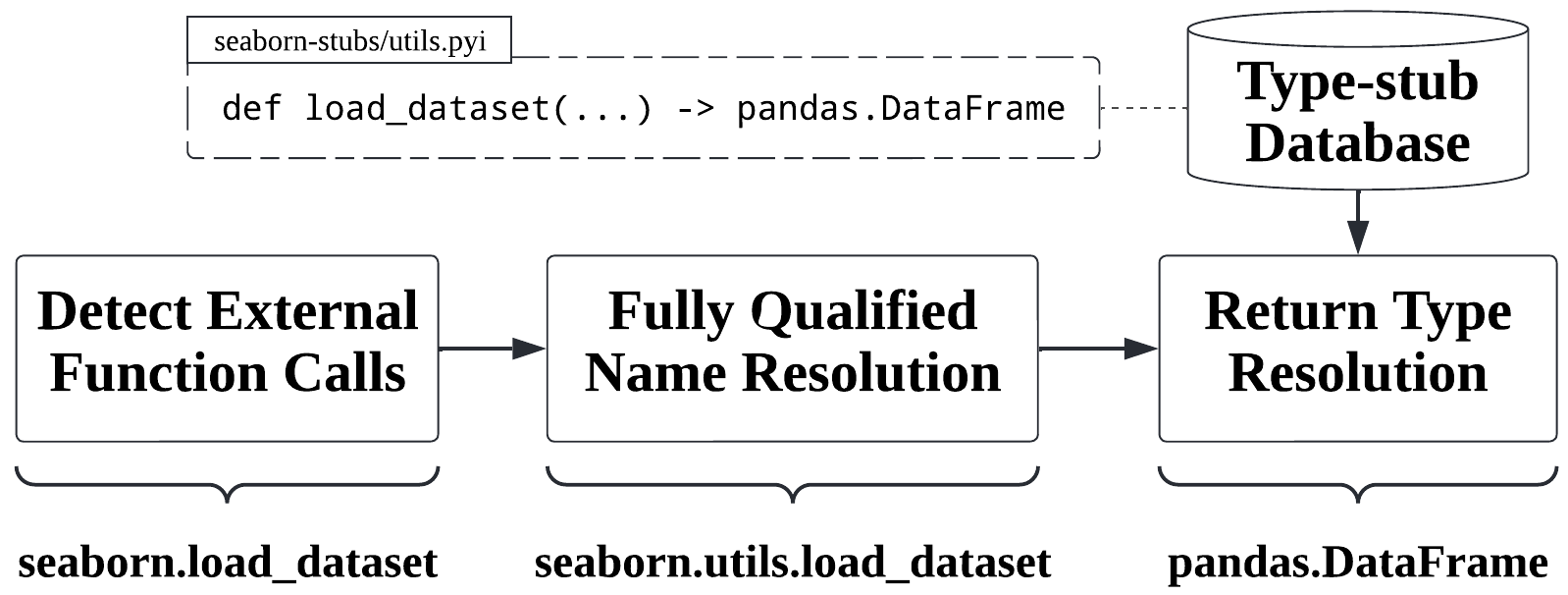}}
	\caption{Workflow of imported library function return-type resolution.}
	\label{fig:ext_workflow}
\end{figure}

We thus instead designed a tool-assisted approximative technique for resolving return-types of function calls to external libraries. 
Figure~\ref{fig:ext_workflow} shows \toolname{}'s approach for return-type approximation.
First, we created a database of stub files for popular ML libraries such as \emph{Keras}, \emph{Numpy}, \emph{Pandas}, etc.
Stub files contain type hints defined relative to the original \python{} source code and stored as \textit{.pyi} file.
To build the database, we first created scaffolding .pyi files for all ML libraries we selected.
This was followed by a manual inspection of function documentation and in some instances, confirmation by manual function execution to create type annotations for individual function calls.
We note that this is still a work in progress and does not yet cover the entire source code of all the ML libraries that we selected.
We intend to fully automate type-stub generation in the future utilizing type-checking systems such as pytype which currently output conservative results that are not usable for our use-case \citep{guoGeneratingPythonType2024}.
However, no accurate and maintained type-inference implementations for \python currently exists.

As shown on the bottom left of Figure \ref{fig:ext_workflow}, additional steps are required to make return-type resolution work for Python.
We took for granted that \texttt{sns.load\_dataset()} resolves to \texttt{seaborn.utils.load\_dataset}. But looking at the example in Figure~\ref{motivatingnb} at (C2,2), this fully qualified function name is not at all apparent.
\toolname thus must implement two additional steps that resolve application-side function calls to their fully qualified names.
First, the external function call in the notebook is resolved based on the import information and EAG.
For instance, consider location (C1,2) in our motivating example.
Here, \texttt{seaborn} is imported with alias \texttt{sns} and therefore the function call is resolved as \texttt{seaborn.load\_dataset}, as shown at the bottom left of Figure~\ref{fig:ext_workflow}.
But in Python, top-level modules can access function definitions in submodules by transitive imports, mapping full path API names to shorter names.
For instance, \texttt{seaborn} exports functions from submodules (\textit{utils.py} here) that actually implement the function. Fortunately, given the fact, that the module \texttt{seaborn} has now been determined,
\toolname can next perform a \emph{dynamic} fully qualified name resolution using the builtin \python{} reflection mechanism \texttt{inspect}, and dynamic execution using the function \texttt{eval} on that module.
During startup time, \toolname{} imports a selected set of popular ML libraries into memory.
Then, during analysis, the \texttt{eval} function is used to dynamically evaluate strings as \python expressions.
In our motivating example, a reference to the function returned by: \texttt{eval(`seaborn.load\_dataset')} is evaluated and stored.
Note that the function \texttt{load\_dataset} is not called---only a reference to the function is dynamically created.
Further, this reference is examined using the builtin \texttt{inspect} module, which can retrieve information about live \python{} objects. 
\toolname uses it to fetch the location of the function's definition in the source code, i.e., the fully qualified name.

\subsubsection{Type Information}
Additionally, \toolname gathers type data for variables found within the program and incorporates this information into the EAG.
This process involves considering types from  both external sources, identified via return-type resolution and local variables.
To facilitate this, we augment PyCG's fixed point algorithm to consider types of variables.

When a variable holds a literal value (such as \texttt{str}, \texttt{int}), \python's built-in reflection mechanism is used to infer the type of the variable.
On the other hand, when the variable is an instance of a user-defined class in the program, the EAG is referred to determine the type of the variable.
This is especially useful when \toolname needs information about variable types to perform heuristic pattern matching on the source code (see Section~\ref{subs:jnannotation}).
Moreover, we posit that having access to flow-sensitive type information of variables can be useful for a broader analysis use case.

\subsection{Flow-sensitive Callsite Extraction}
\label{subs:flowsens}
The EAG generated in the previous step is used to construct a flow-sensitive CG using PyCG's CG construction algorithm.
Wherein, the intermediate representation of the program is iterated while looking for callable objects based on the EAG and adding it to the CG.
Then, the callsites are mapped according to the location of their definition in the notebook.
This is achieved by mapping the line numbers of the \python script with the notebook during conversion.

In addition, note that when a user-defined function that is defined elsewhere in the notebook, say \texttt{x()}, is called from a code cell, any other function called from inside \texttt{x()} is also added as originating from that particular location in the notebook, i.e., the transitive closure of the CG.
This step is needed to ensure that \toolname can annotate code cells that are only calling functions defined in some other code cell.

\subsection{Taxonomy Classifier}
\label{sec:taxonomy_classifier}

Within the Python environment, there exists a vast array of APIs tailored for diverse ML tasks.
These APIs can be classified based on the specific ML operation they perform.
In \toolname, the callsite information obtained from the earlier stage assists in categorizing code cells by the API calls they contain.
In our previous work \citep{venkateshEnhancingComprehensionNavigation2023}, \toolname relied on a manually curated database of ML APIs to ML operations.
However, this is not maintainable considering the fast development cycles of ML libraries. 
Therefore, to aid this process, we developed a machine learning-based classifier for ML taxonomies.
This classifier categorizes ML APIs into distinct ML operations using their documentation strings (doc strings).
In this section, we discuss the process we followed for training this classifier model.

\subsubsection{Dataset Preparation}
In the process of constructing a supervised machine learning model, the initial requisite is a well-labeled dataset. 
Given the absence of any publicly accessible datasets that maps API calls with ML operations, our first step was to establish such a dataset.
To accomplish this, we used Kaggle, a platform that offers a wide range of resources such as datasets and notebooks.
The ML notebooks on Kaggle serve as a valuable resource for practical code implementations with API calls from diverse libraries.
We retrieved 6,698 notebooks from Kaggle's top six competitions, leveraging its API.
The rationale behind selecting these specific competitions was their widespread popularity based on the number of submissions and participating teams.
A breakdown of these competitions and notebooks is listed in Table \ref{table:kaggle_overview}.

We used \toolname to analyze the selected notebooks.
Our primary objective was to extract the fully qualified names associated with all API calls, as well as their corresponding doc strings. 
This data forms the foundational basis for our dataset.

\begin{table}[]
	\renewcommand{\arraystretch}{1.2}
	\centering
	\caption{Overview of Kaggle competitions and notebooks}
	\label{table:kaggle_overview}
	\begin{tabular}{lrr}
		\toprule
		\textbf{Kaggle Competition} & {\textbf{Teams}} & {\textbf{Notebooks}} \\
		\midrule
		Predict Future Sales & 15,656 & 833 \\
		Titanic - Machine Learning From Disaster & 14,560 & 1,089 \\
		Santander Customer Transaction Prediction & 8,751 & 907 \\
		Home Credit Default Risk & 7,176 & 900 \\
		IEEE-CIS Fraud Detection & 6,351 & 820 \\
		M5 Forecasting - Accuracy & 5,558 & 608 \\
		\midrule
		\textbf{Total Notebooks} &  & 6,698 \\
		\bottomrule
	\end{tabular}
\end{table}

\textbf{Empirical Analysis.}
\toolname identified a total of 141,657 API calls from 12 different ML libraries.
After removing the duplicates, we were left with 2,553 unique API calls.
The breakdown of this is listed in Table \ref{table:mostusedlibraries}.

\textbf{Data Annotation.}
To ensure the accuracy and reliability of our annotations, we engaged six experts specializing in machine learning.
These experts were identified through our professional network on LinkedIn.
Given the vastness of the dataset, and in consideration of the potential for annotator fatigue, we opted to select a subset comprising 400 APIs from various libraries. 
This strategic selection was made to ensure a comprehensive coverage across different functionalities while at the same time optimizing the use of our resources.

The chosen APIs are representative of diverse use cases, ensuring a holistic understanding of the domain. 
For example, libraries such as \textit{Matplotlib}, \textit{Seaborn}, and \textit{Plotly} provide an extensive array of APIs dedicated to plotting functionalities. 
On the other hand, \textit{Pandas} and \textit{NumPy} are predominantly known for their data manipulation capabilities.
Similarly, the \textit{Sklearn}, \textit{Keras}, and \textit{Torch} libraries are renowned for their APIs that facilitate model building in ML. 
This structured approach in API selection ensured that our annotations captured the broad spectrum of functionalities inherent in the ML domain.

We developed an annotation tool designed to display the API together with its associated doc string.
This feature was implemented to aid annotators in efficiently identifying and selecting the relevant machine learning operation, subsequently facilitating the exportation of results. 
It is important to note that every API was reviewed by multiple annotators to ensure accuracy. 
To assess the consistency of these annotations, we calculated the inter-annotator agreement and obtained a Cohen's kappa coefficient \citep{cohenCoefficientAgreementNominal1960} of 0.80. 
This score indicates a substantial level of agreement among the annotators.

\begin{table}[]
	\renewcommand{\arraystretch}{1.2}
	\centering
	\caption{Library API utilization overview}
	\label{table:mostusedlibraries}
	\begin{tabular}{lrr}
		\toprule
		\textbf{Library} & {\textbf{Total API calls}} & {\textbf{Unique API calls}} \\
		\midrule
		Pandas       & 61,949 & 393 \\
		Sklearn      & 27,314 & 571 \\
		Matplotlib   & 16,762 & 224 \\
		Numpy        & 12,355 & 273 \\
		Seaborn      & 6,531  & 64  \\
		Keras        & 6,136  & 239 \\
		LightGBM     & 3,100  & 34  \\
		XGBoost      & 2,321  & 37  \\
		Tensorflow   & 1,664  & 345 \\
		Plotly       & 1,482  & 98  \\
		Torch        & 1,449  & 188 \\
		Statsmodels  & 594   & 87  \\
		\midrule
		\textbf{Total}& 141,657 & 2,553 \\
		\bottomrule
	\end{tabular}
\end{table}

\subsubsection{Data Preprocessing}

Doc strings often contain special characters, additional reference texts, and example content that can make ML pattern recognition challenging.
To tackle this, we use Natural Language Processing (NLP) methods to preprocess these strings.
Note that, during the annotation process, the unmodified doc strings were presented to the annotators. 

The steps involved are outlined as follows:
\begin{itemize}
	\item \textbf{Data Cleaning:} removal of LaTeX and markdown formatting strings. Further, we eliminate version numbers, statements indicating deprecation, URLs, punctuation, and other special characters to ensure a cleaner dataset.
	\item \textbf{Stop Word Removal:} commonly used words such as ``the'', ``and'', and ``is'', are removed.
	\item \textbf{Lemmatization:} applied to reduce words to their base form.
\end{itemize}

\begin{table}
	\centering
	\renewcommand{\arraystretch}{1.2}
	\caption{Performance of classifiers for different text vectorization techniques}
	\label{table:performance_metrics}
	\begin{tabular}{l|ccc|ccc|ccc}
		\multicolumn{10}{r}{\footnotesize  \textbf{A:} Accuracy, \textbf{P:} Precision, \textbf{R:} Recall}\\
		\toprule
		{\multirow{2}{*}{\textbf{Classifier}}} & \multicolumn{3}{c}{\textbf{TF-IDF}} & \multicolumn{3}{c}{\textbf{Word2Vec}} & \multicolumn{3}{c}{\textbf{CountVec}} \\
		\cmidrule(lr){2-4}\cmidrule(lr){5-7} \cmidrule(lr){8-10}
		& A. & P. & R. & A. & P. & R. & A. & P. & R. \\
		\midrule
		LogisticRegression & 0.86 & 1.0 & 0.86 & 0.60 & 0.95 & 0.59 & 0.89 & 0.98 & 0.89 \\
		RandomForest & 0.84 & 0.99 & 0.86 & 0.80 & 0.98 & 0.80 & 0.88 & 0.99 & 0.88 \\
		DecisionTree & 0.76 & 0.89 & 0.82 & 0.57 & 0.74 & 0.68 & 0.69 & 0.81 & 0.88 \\
		GaussianNB & 0.83 & 0.92 & 0.86 & 0.72 & 0.81 & 0.83 & 0.75 & 0.84 & 0.88 \\
		\rowcolor{black!5}\textbf{SVM} & \textbf{0.94} & \textbf{1.0} & \textbf{0.94} & 0.83 & 0.89 & 0.91 & 0.77 & 0.93 & 0.80 \\
		GradientBoosting & 0.81 & 0.96 & 0.82 & 0.79 & 0.92 & 0.81 & 0.85 & 0.95 & 0.89 \\
		\bottomrule
		
	\end{tabular}
\end{table}

\subsubsection{Model Training}

We used the finalized dataset with a train-test split of 80\%-20\% to train and test a series of multi-label classification techniques. 
The nature of our classification problem is multi-label because a single API can correspond to multiple ML Operations. 
For instance, the API \mbox{\texttt{numpy.ndarray.reshape}} can be simultaneously categorized under both ``Data Preparation and Exploration'' and ``Feature Engineering''.

As ML Models operate on numerical data, we first convert doc strings into a numerical format, in a process termed as text vectorization.
In this research, we explored three prominent text vectorization methodologies: Word2Vec, TF-IDF, and CountVectorizer. 
With data from these techniques, we subsequently trained a range of ML models listed as follows: Logistic Regression, Random Forest, Decision Tree, GaussianNB, Support Vector Machines (SVM), and Gradient Boosting. 
Note that, driven by the constraints posed by our small dataset size, we abstained from incorporating neural networks in our experimentation.

\textbf{Model Selection.} 
The performance of various classifier models is shown in Table \ref{table:performance_metrics}.
In comparison with other classifiers and across vectorization methods, SVM stands out particularly with TF-IDF, showcasing a balance of accuracy, precision, and recall.
Therefore, within \toolname we integrated the SVM classifier with TF-IDF vectorization for classifying API calls.

\subsection{Jupyter Notebook Annotation}
\label{subs:jnannotation}

The goal of \toolname{} is to aid ML practitioners in easily navigating and comprehending undocumented \jnb{}s.
To this end, the callsite information output by \toolname's analyzer
is used to add helpful information to the notebook.

\subsubsection{Pattern Matching}
Notebooks can contain code cells that perform ML operations without explicit function calls, but rather, use other \python constructs that alter objects.
In absence of a function call, \toolname resorts to AST based pattern matching to identify ML operations.

The process is as follows: Initially, \toolname identifies code cells devoid of any function calls by analyzing the flow-sensitive CG constructed in the preceding phase of the analysis.
Subsequently, it traverses the AST of these code cells to detect the presence of specific patterns outlined in Table \ref{table:usage_patterns}.
However, given that ASTs lack type information, \toolname consults the EAG using line number and node identifier from the AST node, \texttt{Name}, to retrieve type information for AST elements identified within the code cell. 
Finally, if a pattern aligns with one of the supported patterns by \toolname, the corresponding code cell is annotated accordingly.

For instance, consider the first pattern in Table~\ref{table:usage_patterns} that represents a \emph{Feature Engineering} operation, i.e., $\texttt{df{[}`xy'{]} = df.x * df.y}$.
Here, a new column \emph{xy} is being created in the Dataframe object \texttt{df} by multiplying columns \emph{x} and \emph{y}.
In this specific case, \toolname checks if both sides of the \texttt{BinOp} AST node for the binary operator `$*$' are indeed Dataframe accesses.
That is, if \texttt{df.x} and \texttt{df.y}, are Dataframe accesses.
Just by looking at the AST, \toolname cannot ascertain the type of the variable \texttt{df}, consequently, it retrieves the type information from the EAG to check if \texttt{df} is a Dataframe. 
Then, 
From this, \toolname concludes that this statement is a \emph{Feature Engineering} operation.
Table \ref{table:usage_patterns} lists the Dataframe access patterns that \toolname currently supports.

\begin{table}[t]
	\renewcommand{\arraystretch}{1.2}
	\centering
	\caption{Mapping of Dataframe usage patterns to ML operations}	
	\label{table:usage_patterns}	
	
	\begin{tabularx}{.8\textwidth}{lXX}
		\toprule
		\textbf{ID} & \textbf{Pattern} & \textbf{ML Operation} \\
		\midrule
		1 & \texttt{df{[}`xy'{]} = df.x * df.y} & Feature Engineering \\
		\rowcolor{black!5}2 & \texttt{df.x = 1} & Feature Transformation \newline Data Preparation \\
		3 & \texttt{df.x{[}df.x == 1{]} = 1} & Feature Transformation \newline Data Preparation \\
		\rowcolor{black!5}4 & \texttt{x = df.x{[}{[}`f1', `f2'{]}{]}} & Feature Selection \\
		5 & \texttt{print(df{[}0:20{]})} & Exploratory Data Analysis \\
		\rowcolor{black!5}6 & \texttt{for i in df:} & Data Preparation \\
		7 & \texttt{len(df)} & Exploratory Data Analysis \\
		\rowcolor{black!5}8 & \texttt{df.shape} & Exploratory Data Analysis \\
		\bottomrule
	\end{tabularx}
\end{table}

\begin{figure}[]
	\centering

	\subfloat[Index of ML operations for our motivational example. 
	\raisebox{.5pt}{\textcircled{\raisebox{-.9pt} {1}}}	ML operation category ``Model Training" is expanded to view all code cells that are performing model training operations.
	\raisebox{.5pt}{\textcircled{\raisebox{-.9pt} {2}}} Cell \# 5 is expanded to view all function calls in the cell.
	\raisebox{.5pt}{\textcircled{\raisebox{-.9pt} {3}}} Fully qualified function names are displayed.
	\raisebox{.5pt}{\textcircled{\raisebox{-.9pt} {4}}} Expanded view showing the arguments used and its docstring.]{\includegraphics[width=.95\linewidth]{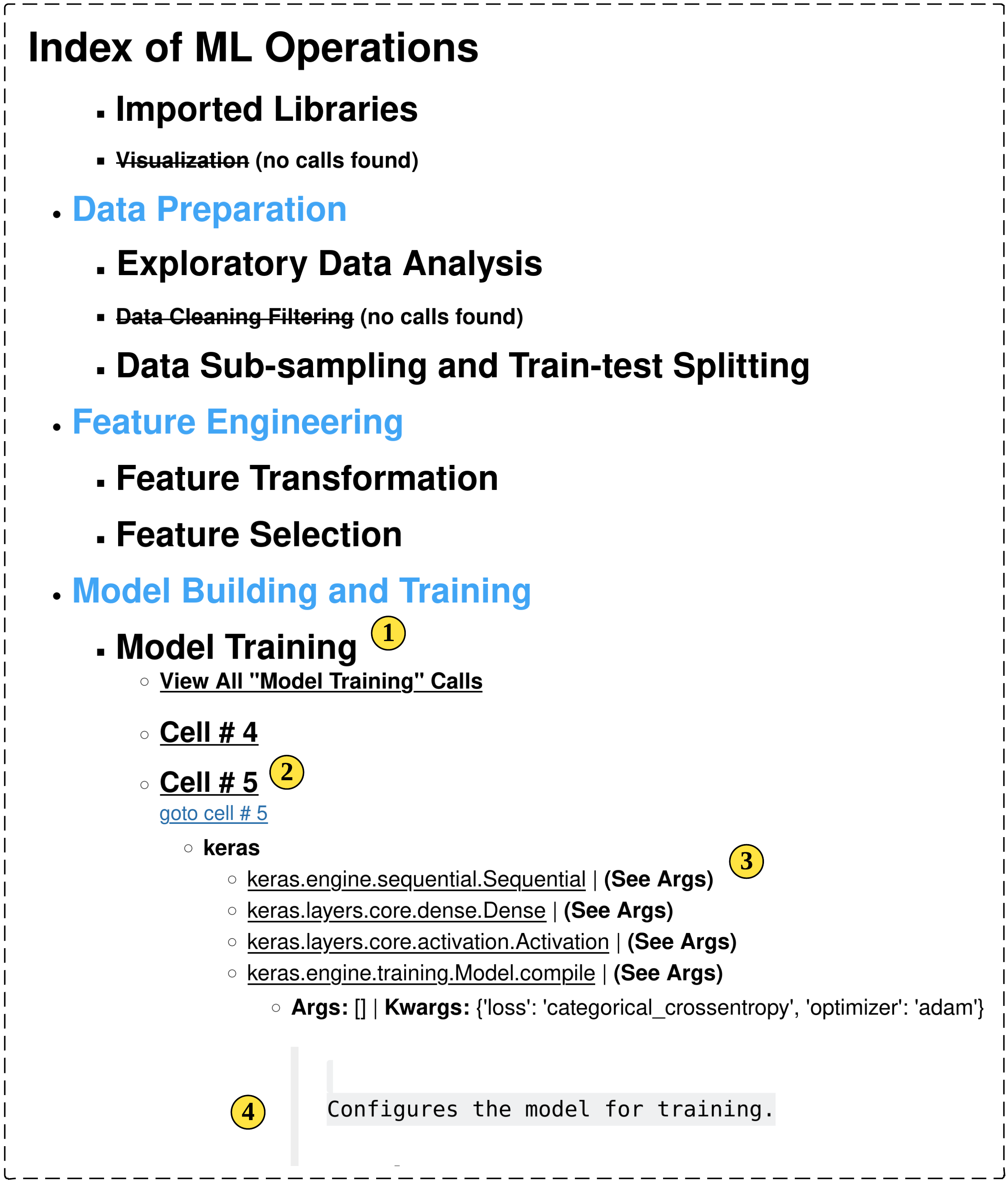}
	\label{fig:index_ml_ops}}

	\subfloat[Annotated version of cell \# 3 in our motivational example.]{\includegraphics[width=.95\linewidth]{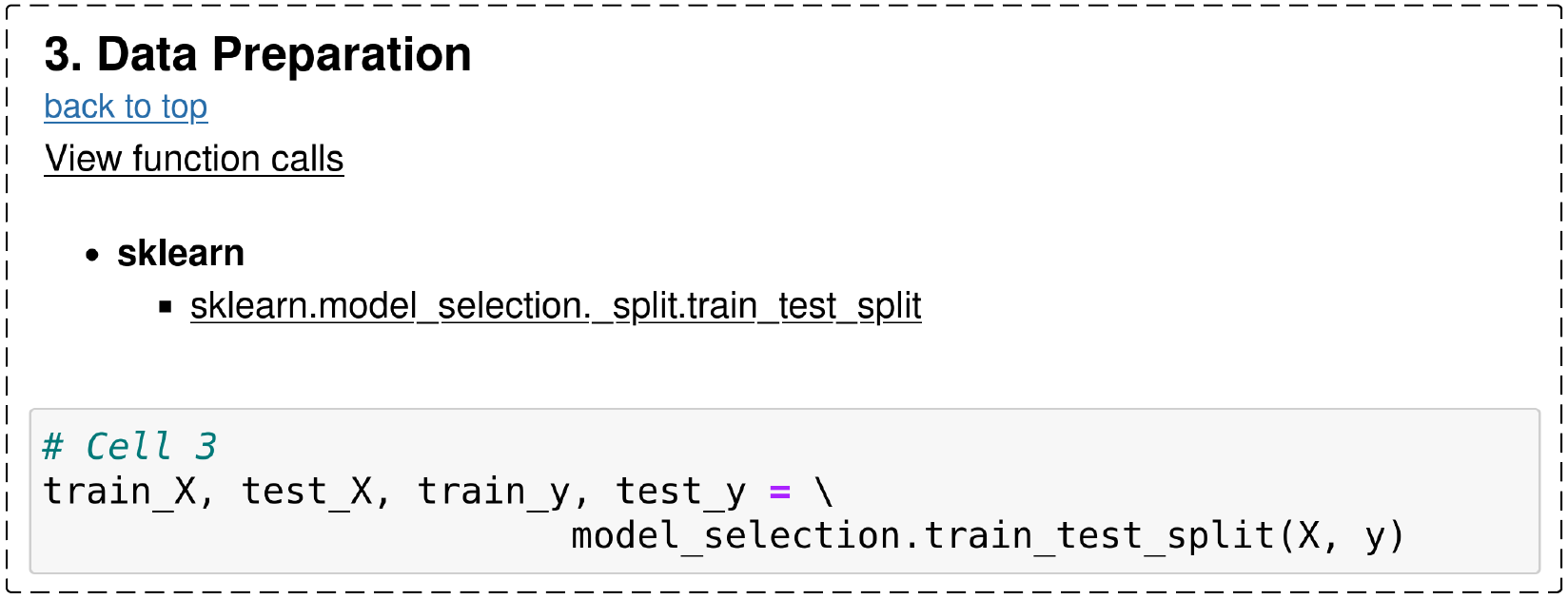}
	\label{fig:codecell_annotated}}

\caption{Snapshots of the output notebook generated by \toolname for our motivational example.} 
\end{figure}

\subsubsection{Text Annotation Generation}
Based on this classification and pattern matching, the following annotations are added to the notebook: (1) Index of ML Operations, (2) Code cell headers, and (3) Table of contents.

\textbf{1) Index of ML Operations:} 
The index provides a clickable and nested list of all function calls in the notebook classified according to the taxonomy of ML operations shown in Figure~\ref{t:ml_operations}.
Figure \ref{fig:index_ml_ops} shows the index of ML operations generated for our motivating example.
The index is displayed on top of the notebook using \toolname's \jnb{} plugin.
If no functions are found for a particular ML operation category, the category is displayed struck out.
Each ML operation category and cell list can be expanded or collapsed as required.
Function calls are organized based on the library as seen in the figure.
Additionally, different areas of the notebook are hyperlinked, this makes it easy for the user to explore the notebook back-and-forth.
For instance, cell~5 can be quickly visited by pressing \textit{``goto cell \# 5''} and back to the index again by pressing \textit{``back to top''}.

\textbf{2) Code cell headers:}
High-level ML operation categories from the taxonomy are added as headers for individual code cells.
Note that when code cells contain ML operations from more than one category, all of these are added to the header.
Figure \ref{fig:codecell_annotated} shows the annotated version of code cell~3 from our motivating example (Figure \ref{motivatingnb}).
The headers can be further extended to see all the functions used in the following code cell, along with the docstrings that were fetched during analysis time from the source code.

\textbf{3) Table of contents:}
Code cell headers are attached with anchors that allow in-page navigation.
Using this information, the table of contents combines the headers of all code cells and adds an anchor-link to each entry.
This simplifies access to relevant sections of the notebook based on the taxonomy.

\section{Evaluation}
\label{sec:evaluation}

We evaluated \toolname to answer the following four research questions:

\begin{itemize}[leftmargin=1cm]
	\itemsep.5em
	\item[\textbf{\textit{RQ1:}}] \textit{Does HeaderGen improve comprehension and navigation of undocumented Jupyter Notebooks?}
	
	\item[\textbf{\textit{RQ2:}}] \textit{How accurate is HeaderGen's callsite recognition?}
	
	\item[\textit{\textbf{RQ3:}}] \textit{How accurately can HeaderGen classify code cells using callsites?}

	\item[\textit{\textbf{RQ4:}}] \textit{To what extent does type information contribute to the accuracy of HeaderGen?}

	\item[\textit{\textbf{RQ5:}}] \textit{How does HeaderGen compare to other tools?}
	
\end{itemize}

We first describe the benchmarks we developed for evaluating \toolname in Section \ref{sec:benchmarks}, and then examine the research questions.

\section{Benchmarks}
\label{sec:benchmarks}
We evaluate \toolname by building four benchmarks: 
(1) a micro-benchmark containing 121 \jnb{}s,
(2) a real-world benchmark containing 15 \jnb{}s from Kaggle,
(3) an extended real-world benchmark containing 15 \jnb{}s from the wild, and
(4) \typeevalpy, a framework for benchmarking type inference in \python.
We discuss these benchmarks in the subsequent sections.

\subsection{Jupyter Notebook Micro-benchmark}
We evaluate \toolname by adopting the benchmark created by \cite{salisPyCGPracticalCall2021a} as part of PyCG.
PyCG's benchmark does not have specific challenges targeting flow-sensitive analysis, and the benchmark contains ground truth only for flow-insensitive call-graphs.
Yet, to evaluate \toolname's analysis, flow-sensitive callsite information is required, i.e., information about function calls associated with line numbers.
To address this, we first converted \python{} scripts from PyCG's benchmark into \jnb{}s, and then created ground truth by manually mapping callsites to line numbers.
Furthermore, we created eight new test cases that have specific challenges to flow-sensitivity.

\subsection{Real-world Benchmark}
\label{subsec:rwb}
To assess \toolname in real-world scenarios, we tested for precision and recall on 15 \jnb{}s from Kaggle, a community where ML practitioners come together to create and share ML based solutions written in \jnb{}s.
The platform hosts open competitions where data scientists around the world compete against each other to build the best solution.
Kaggle encourages beginners to learn from experts in the field by making their submissions public.
However, these notebooks often lack documentation.
We found that 99 of the top 500 notebooks submitted to the most popular competition on Kaggle contained no markdown cell.
Therefore, we base our real-world benchmark on these undocumented notebooks which are still being viewed by many (cf. Table \ref{table:realworldbench_notebooks}).

We selected notebooks from three different and most popular competitions on Kaggle based on the number of submissions to encourage variation in the benchmark:
(1) Titanic - Machine Learning from Disaster,
(2) Predict Future Sales, and
(3) Santander Customer Transaction Prediction.
We downloaded the top 30 notebooks according to votes for each competition with the search term ``Keras'',
since Keras\footnote{\url{https://keras.io/}} is a popular ML library among novices.
We used the Kaggle API to search and download notebooks.
All 30 notebooks from each competition were further filtered to target those without any markdown cells.
Finally, we selected the top five most viewed notebooks from each competition.
The selected notebooks in our benchmark are listed in Table~\ref{table:realworldbench_results}.
These notebooks have a median of 20 code cells, compared to 13 cells that are found in real-world notebooks as reported by \cite{pimentelLargeScaleStudyQuality2019}.
Note that these undocumented notebooks still have 240 upvotes and 17,687 views as of Octorber 2023. 

The callsites ground truth was created manually by inspecting code cells in each notebook, and listing the fully qualified names of all function calls.
Notebooks were executed cell-by-cell and dynamically analyzed using \python{}'s reflection module \texttt{inspect} to gather the fully qualified names.
Multiple iterations were carried out to avoid errors in the ground truth.

Further, the ground truth for headers were created using experts.
15 notebooks from the benchmark were divided and assigned to four data scientists working in the industry for manual annotation of each code cell. 
Notebooks were distributed such that each notebook was seen by at least two reviewers.
Based on the taxonomy of ML operations, each annotator inspected and classified each code cell into relevant categories.
The inter-rater reliability score, as measured by Cohen's kappa coefficient \citep{cohenCoefficientAgreementNominal1960}, was improved by conducting follow-up interviews with all four reviewers.
Finally, a score of 0.89 was achieved, which signals an almost perfect agreement.

\begin{table}[]
	\renewcommand{\arraystretch}{1.2}
	\centering
	\caption{Details of notebooks included in the real-world benchmark evaluation}	
	\label{table:realworldbench_notebooks}	
	\begin{tabularx}{.9\textwidth}{clcc}
		\toprule
		\multirow{2}{*}{\textbf{ID}} & \multicolumn{1}{c}{\textbf{Name}}                  & \multicolumn{1}{c}{\multirow{2}{*}{\textbf{Votes}}} & \multicolumn{1}{c}{\multirow{2}{*}{\textbf{Views}}} \\
		& \multicolumn{1}{c}{\textbf{}}                      & \multicolumn{1}{c}{}                                & \multicolumn{1}{c}{}                                \\ \hline
		1                            & bulentsiyah/keras-deep-learning-to-solve-titanic   & 69                                                  & 2,244                                               \\
		2                            & hongdnghuy/relu-sigmoid                            & 13                                                  & 743                                                 \\
		3                            & vaidicjain/titanic-easy-deeplearning-acc-78        & 9                                                   & 489                                                 \\
		4                            & tanvikurade/complete-analysis-of-titanic           & 19                                                  & 326                                                 \\
		5                            & alexanderbader/mytitanic                           & 10                                                  & 113                                                  \\ \hline
		6                            & econdata/predicting-future-sales-with-lstm         & 7                                                   & 3,363                                               \\
		7                            & lhavanya/predict-future-sales                      & 3                                                   & 500                                                 \\
		8                            & elvinagammed/stacked-lstm-top-5-4-mae              & 9                                                   & 523                                                 \\
		9                            & ashishkapasiya/prediction-future-sales-with-keras  & 4                                                   & 525                                                 \\
		10                           & the0electronic0guy/keras-begineer-friendly         & 12                                                  & 290                                                 \\ \hline
		11                           & higepon/starter-keras-simple-nn-kfold-cv           & 20                                                  & 4,387                                               \\
		12                           & vishesh17/keras-nn-with-scaling-and-regularization & 32                                                  & 3,205                                               \\
		13                           & christofhenkel/nn-with-magic-augmentation          & 21                                                  & 1,573                                               \\
		14                           & naivelamb/multibranch-nn-baseline-magic            & 10                                                  & 652                                                 \\
		15                           & miklgr500/nn-embedding                             & 10                                                  & 585                                                 \\ \hline
		& \multicolumn{1}{r}{\textbf{Total}}                 & \textbf{248}                                        & \textbf{19,518} \\	\bottomrule

	\end{tabularx}	
	
\end{table}

\subsection{Extended Real-world Benchmark}

Building upon our initial publication \citep{venkateshEnhancingComprehensionNavigation2023}, we have extended our benchmark suite by incorporating an additional 15 Jupyter notebooks.
This extension aims to further assess the utility of \toolname, across a broader spectrum of ML notebooks found in real-world contexts.
The process of annotating notebooks with metadata pertaining to callsites and headers presents significant resource demands.
Fortunately, \cite{ramasamyWorkflowAnalysisData2022} recently introduced the DASWOW dataset.
This dataset comprises 470 notebooks annotated by domain experts, albeit using a different classification system that does not directly correlate with the taxonomy used by \toolname.

Moreover, to evaluate \toolname's accuracy, we also require ground truth of the callsites within these notebooks.
Given the resource requirements of generating ground truth for the entirety of the notebooks, particularly regarding callsites, we narrowed our focus to a set of 15 notebooks.
To this end, we first selected notebooks devoid of markdown cells from the DASWOW dataset, resulting in 124 notebooks.
Subsequently, we prioritized notebooks with more number of code cells, selecting the top 15 notebooks. This collection contains a median of 27 code cells.

The creation of ground truth for headers involved aligning the DASWOW taxonomy with our own, leveraging the partial mappings provided by \cite{ramasamyWorkflowAnalysisData2022} in their work.
This was followed by a manual verification of header annotations by the first author, to ensure accuracy and to add any missing labels.
Furthermore, the callsites for notebooks were determined through a manual review process, adhering to the methodology described in the preceding Section \ref{subsec:rwb}.

\subsection{TypeEvalPy: Framework for Type Inference Benchmarking}
\label{section:typeevalpy}
In recent years, the development and publication of type inference tools for Python have experienced a surge in both academic research and the open-source community.
However, to date, no attempt has been made to construct a comprehensive benchmark dedicated to evaluating type inference tools for Python.
The majority of existing research evaluates the accuracy of type inference tools using real-world benchmark datasets, for instance \emph{Type4Py}~\citep{type4py} and \emph{HiTyper}~\citep{HiTyper}.
On the other hand, open-source tools are evaluated using specifically-designed test cases.

Such evaluation methods, however, come with their set of challenges:
\begin{itemize}
	\item Different studies may use different datasets, making it difficult to directly compare and understand the strengths and weaknesses of each tool.
	\item Real-world datasets occasionally contain inaccurate type annotations.
	\item Many evaluations give a general score, missing detailed insights into specific challenges, such as handling different language features.
\end{itemize}

To address this gap, we propose \typeevalpy, a comprehensive micro-benchmark suite aimed at evaluating type inference in Python, encompassing the diverse and complex features of Python.
Our micro-benchmark comprises 154 code snippets, each focusing on distinct features of the Python language, such as, generators and decorators.
The first two authors conducted exhaustive ground-truth type annotations for each snippet, subjecting each to multiple reviews during the developmental stages to ensure accuracy. A foundational objective of \typeevalpy is to ensure reproducibility and extensibility to any type inference tool. 
To realize this, we incorporated Docker and Python boilerplate code, with uniformly implemented functions enabling easy integration and execution of any type inference tool on the micro-benchmark.
To ensure reproducibility of results, \typeevalpy is highly automated, allowing the evaluation of the benchmark across all tools with the execution of a single command.
Additionally, using the provided boilerplate, we have implemented containers compatible with \typeevalpy for six renowned Python type inference tools: \emph{\toolname}, \emph{Jedi}, \emph{Pyright}, \emph{HiTyper}, \emph{Scalpel}, and \emph{Type4Py}.

\begin{figure}[]
	\centering
	\includegraphics[width=\linewidth]{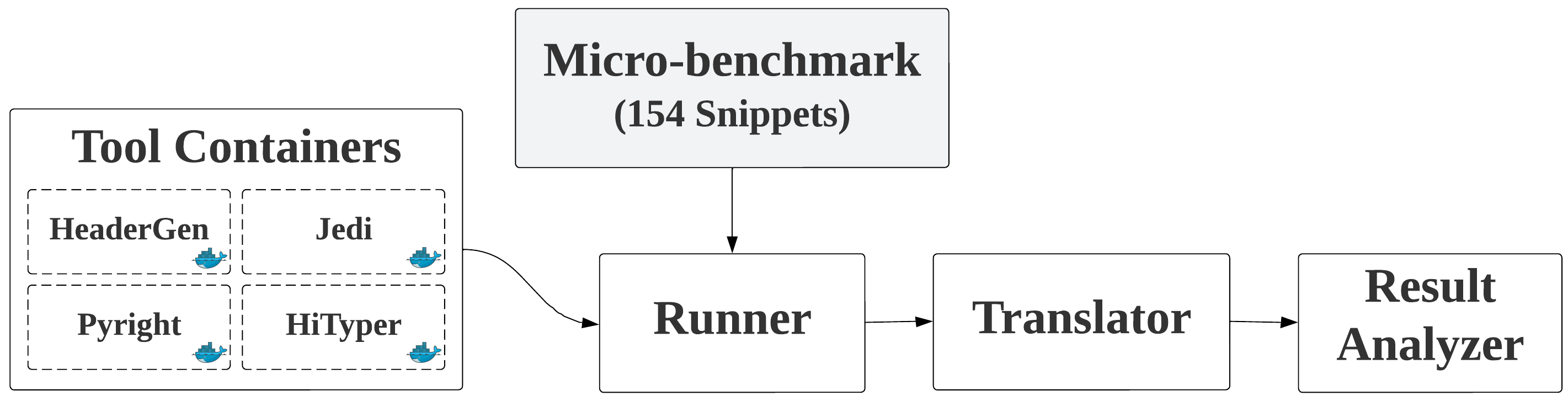}
	\caption{Overview of benchmarking workflow in \typeevalpy}
	\label{fig:typeevalpy_overview}
\end{figure}

The workflow of \textit{\typeevalpy} is outlined in Figure \ref{fig:typeevalpy_overview}. The \texttt{runner} performs tests on each type inference tool in \textit{\typeevalpy} using the micro-benchmark. After completion, the \texttt{translator} converts results into a standard format defined in the \textit{\typeevalpy} framework.
Note that each tool implements its own translation logic and the \texttt{translator} orchestrates the conversion.
These files are then analyzed to produce detailed evaluation metrics for type inference by the \texttt{result analyzer}. Further details about these components are discussed in the following sections.

\subsubsection{Micro-benchmark}

The \typeevalpy micro-benchmark encompasses 154 code snippets, systematically organized into 18 categories, reflective of various Python features.
These categories are designed, taking inspiration from the PyCG call-graph benchmark and contains the following categories: 
\emph{args (8)}, \emph{assignments (8)}, \emph{builtins (7)}, \emph{classes (26)}, \emph{decorators (8)}, \emph{dicts (15)}, \emph{direct\_calls (6)}, \emph{dynamic (3)}, \emph{exceptions (2)}, \emph{external (7)}, \emph{functions (9)}, \emph{generators (6)}, \emph{imports (14)}, \emph{kwargs (4)}, \emph{lambdas (6)}, \emph{lists (10)}, \emph{mro (7)}, \emph{returns (8)}.

The corresponding ground truth contains a total of 845 type annotations from 154 code snippets, each of which contains the name of the entity, line number, and column offset, supplemented by the category of the type.
Type annotations are cataloged into three distinct categories: 1) Function return~(FR) type, 2) Function parameter~(FP) type, and 3) Local variable (LV) type.
The creation of ground truth involved the validation of runtime types through the execution of the program and the use of a debugger.

Several considerations were taken during the design phase of the benchmark:
\begin{enumerate}
	\item Annotations are flow-sensitive, this means they are connected to the location or line number of the entity.
	\item The type of function return, function parameters, and local variables can be different based on the calling context within a function definition. The type annotations are context-\emph{in}sensitive, though, thus enumerate the union of all types in all concrete usage contexts of the application at hand.
	\item Annotations use generic types. For example, a list of integers is noted as just a \texttt{List}, not \texttt{List[int]}. However, individual elements of lists and dicts receive their own annotations.
	\item Entities are not labeled with the ``Any'' type. When a function has the potential to return multiple types, efforts are made to include all possible return types within the calling contexts.
	\item If a function does not have a return statement, we specifically annotate it as \texttt{Nonetype}.
\end{enumerate}

\subsubsection{TypeEvalPy Runner}

In the \typeevalpy framework, the role of the \texttt{runner} is pivotal.
It initiates the containerized type inference tools and methodically executes type inference on every snippet included in the micro-benchmark. 
The \texttt{runner} component also orchestrates the translation of the results into the \typeevalpy format by using the \texttt{translator} component.
The \texttt{runner} gathers all the derived results and consolidates them. 
After the translation, the \texttt{runner} subjects the results to the \texttt{result analyzer} to extract meaningful insights. 

The \texttt{translator} and \texttt{result analyzer} components are discussed in the following sections.

\subsubsection{TypeEvalPy Translator}

Various type inference tools present their analysis outcomes in diverse formats.
Therefore, the \typeevalpy framework establishes a common schema to describe types, based on the Scalpel framework~\citep{liScalpelPythonStatic}.
Each tool should implement a \texttt{translator} plugin according to the specifications of the \typeevalpy framework.

Listing~\ref{lst:typeevalpy_format_types} shows all kinds of type annotations within the \typeevalpy schema corresponding to the code snippet shown in Listing \ref{lst:typeevalpy_format_program}.
Below are the categories of type annotations along with their corresponding line numbers as depicted in Listing \ref{lst:typeevalpy_format_types}:

\begin{itemize}
	\item \textbf{Function return:} \texttt{id\_func}: line 2-5
	\item \textbf{Function parameter:} \texttt{arg}: line 8-12
	\item \textbf{Local variables:}	\texttt{x}: line 15-19,	\texttt{result}: line 22-25 and 28-31
\end{itemize}

Note that the annotation for the local variable \texttt{x}, at lines 15-19, incorporates a ``function'' key, serving to signify its definition locality.
Conversely, the annotation for the local variable \texttt{result}, at lines 22-25 and 28-31, omits this key, attributed to its definition at the module level.
Note that the \texttt{result} variable is annotated with flow-sensitivity at two program points.

\begin{minipage}[t]{.5\textwidth}
	\captionsetup[lstlisting]{singlelinecheck=off,justification=raggedright}
	\begin{lstlisting}[language=Python, caption=simple\_code.py, style=pstyle, label=lst:typeevalpy_format_program]
def id_func(arg):
	x = arg
	return x

result = id_func("String")
result = id_func(1)
\end{lstlisting}
\end{minipage}%
\begin{minipage}[t]{.5\textwidth}
	\captionsetup[lstlisting]{singlelinecheck=off}
	\begin{lstlisting}[caption=Types for simple\_code.py, style=pstyle, label=lst:typeevalpy_format_types]
[{
   "file": "simple_code.py",
   "function": "id_func",
   "line_number": 1,
   "type": ["int", "str"]
},
{
   "file": "simple_code.py",
   "function": "id_func",
   "line_number": 1,
   "parameter": "arg",
   "type": ["int", "str"]
},
{
   "file": "simple_code.py",
   "function": "id_func",
   "line_number": 2,
   "type": ["int", "str"],
   "variable": "x"
},
{
   "file": "simple_code.py",
   "line_number": 5,
   "type": ["str"],
   "variable": "result"
},
{
   "file": "simple_code.py",
   "line_number": 6,
   "type": ["int"],
   "variable": "result"
}]
\end{lstlisting}
\end{minipage}

\subsubsection{TypeEvalPy Result Analyzer}

The \texttt{result analyzer} component within the \typeevalpy framework generates a comprehensive set of comparative statistics to evaluate the performance of various tools. 
Specifically, it computes the following metrics:

\begin{itemize}
	\item \textbf{Exact matches:} \textit{How many of the inferred types exactly match the ground truth?}
	\item \textbf{Precision:} \textit{Of the types reported, how many are exactly inferred according to the ground truth?}
	\item \textbf{Recall:} \textit{How many of the actual types are exactly reported by the type inference tool?}
	\item \textbf{Soundness:} \textit{Does the type inference tool exactly identify all possible types specified in the Python code to ensure none are omitted?}
	\item \textbf{Completeness:} \textit{Does the tool accurately report only the types that are present, avoiding any incorrect or extraneous types?}
	\item \textbf{Top-n prediction comparison:} \textit{How do the ML-based tools compare in terms of accuracy when considering their top-n inferred types?}
	\item \textbf{Report of missing types:} \textit{Which types, present in the ground truth, are missed by the tools?}
	\item \textbf{Report of mismatched types:} \textit{Which types reported by the tools do not align with the types present in the ground truth?}
\end{itemize}

	\section{RQ1: Comprehension and Navigation Study}
	\label{sec:userstudy}
	The goal of \toolname{} is to increase comprehension and navigation in undocumented \jnb{}s.
	We therefore conducted a user-study to quantitatively measure the improvements of \toolname over undocumented \jnb{}s.
	
	\subsection{Study Design} 
	The study is aimed at recreating the exploration of notebooks that ML practitioners routinely do.
	The study is designed as a within-subject study where the participants were given two notebooks from our real-world benchmark and asked to complete five comprehension tasks on each notebook one after the other.
	To minimize learning effects, we chose a latin-square design:
	participants were divided into two groups.
	While participants in group-1 were given the undocumented notebook first, followed by the \toolname{} annotated version, participants in group-2 saw the annotated notebook first.
	Each study was conducted in a one-on-one online session lasting about one hour using a video-conferencing tool.
	First, an overview of the study-protocol was presented to the participant including a walk-through of \toolname{}.
	Next, participants were provided access to the remote Jupyter instance along with a questionnaire containing step-wise instructions on how to proceed.
	Before proceeding to the study, participants were instructed to examine an example notebook annotated with \toolname{} in order to get them comfortable with the features.
	The entire session was recorded with the consent of the participant for further analysis.
	Upon completion of the comprehension tasks, participants were asked to fill a likert-scale questionnaire to understand the participant's perception of improvements provided by \toolname{}.
	Finally, participants were asked if they had any general comments about the tool.
	
	\textbf{Comprehension Tasks.}
	We created a set of tasks to simulate typical questions that arise when a data scientist is exploring an unseen notebook.
	The tasks were finalized after discussions with a data-science expert.
	For each task, participants were expected to select the right answers from all the choices given to them.
	Overall, six comprehension tasks were created, as listed in Table~\ref{table:comprehensiontasks}.
	For each notebook given to the participant, five tasks from the table were assigned to them based on the relevance to the notebook.
	
	\begin{table}[t]
	\renewcommand{\arraystretch}{1.2}
	\centering
	\caption{Comprehension tasks}	
	\label{table:comprehensiontasks}
	
	\begin{tabularx}{.95\textwidth}{clX}
		\toprule
		\textbf{Id} & \textbf{Question} \\
		\midrule
		Q1 & What are the deep learning layers used in the model? \\
		Q2 & What are the different data cleaning \& data preparation operations? \\
		Q3 & Which of the following cells are used for model building and model training? \\
		Q4 & Select ML and visualization libraries that are used in the notebook \\
		Q5 & What are the different visualizations used in the notebook? \\
		Q6 & How is the dataset split into test and train subsets? \\
		\bottomrule
	\end{tabularx}
\end{table}
	
	\textbf{Likert-scale Questionnaire.}
	Following the completion of the session, participants were asked to rate the level of agreement to statements about the usefulness of \toolname{}.
	The level of agreement was based on a 5-point Likert scale, where ``1'' is \textit{Strongly disagree} and ``5'' is \textit{Strongly agree}.
	The statements given to the participants are listed in Table \ref{table:likertstatements}.

	\begin{table}[t]
		\renewcommand{\arraystretch}{1.2}
		\centering
		\caption{Statements concerning the perception of usefulness}	
		\label{table:likertstatements}
	 \begin{tabularx}{.95\textwidth}{lX}
		\toprule
		\textbf{Id} & \textbf{Statement}   \\
		\midrule
		S1 & The classification of cells according to ML phases and headers helped me navigate the undocumented notebook. \\
		S2 & The generated list of functions used in the notebook helped me understand the notebook better. \\
		S3 & The header annotations added to the notebook are rather hindering the understanding of the notebook. \\
		S4 & I would install \toolname if it is made available as a plugin. \\
		\bottomrule
	\end{tabularx}
	\end{table}

	\subsection{Participants}
	The study comprised of eight participants.
	Three of them were master students from the computer science department,
	three of them were full-time employees working in the data-science domain, 
	and two of them were computer science researchers.
	Students were recruited by contacting the group leaders in the data-science research department.
	Professional employees were contacted using Linkedin\footnote{\url{https://www.linkedin.com/}} based on their job titles.
	The researchers were contacted based on their publications in common research topics.
	Due to privacy concerns, information of the participants are omitted.
	Participation was voluntary and did not involve monetary incentives.
	
	\subsection{Metrics}
	\begin{description}
		\item[(1) Time:]
		Time taken to complete all five tasks per notebook.

		\item[(2) Accuracy:]
		Inspired by a similar comprehension study by \cite{adeliSupportingCodeComprehension2020a}, the accuracy is measured using F1-score that takes into account both precision and recall.

		\item[(3) Navigability:]
		The perceived navigability based on responses to Likert scale questions.
		
		\item[(4) Usefulness:]
		The perceived usefulness based on responses to Likert scale questions.
		
	\end{description}
	
	\subsection{Results}
\begin{figure}[]
		\centerline{\includegraphics[width=\linewidth]{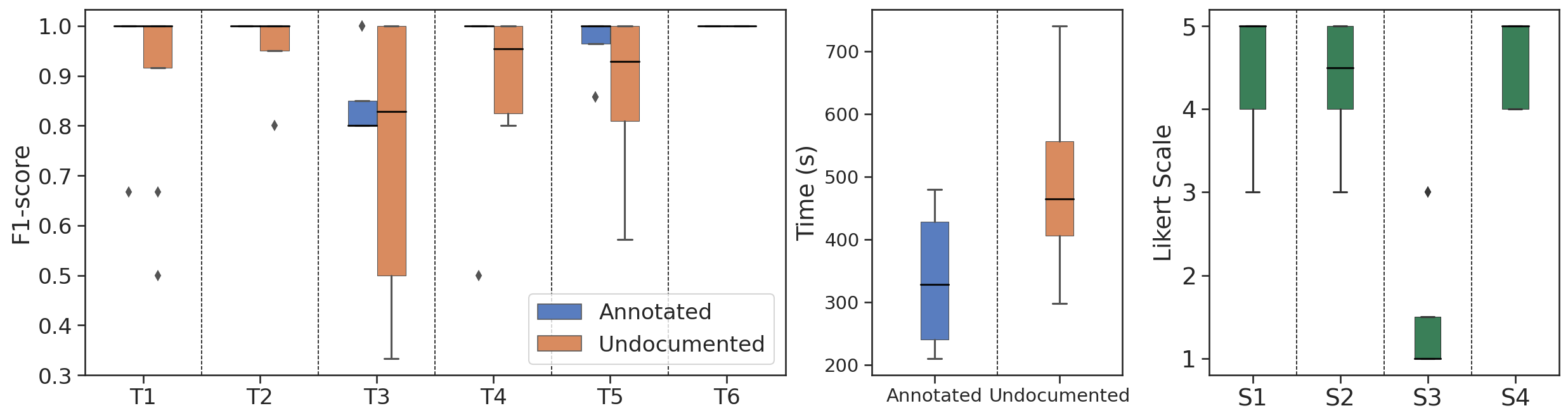}}
		\caption{\textbf{Left:} Box plots of accuracy for participant responses grouped by treatment.
			\textbf{Center:} Box plots of time measurements for two treatments.
			\textbf{Right:} Box plots of responses to likert-questions about perception.}
		\label{fig:combined_graph}
		
	\end{figure}

	The study resulted in 80 ($8 * 5 * 2$) measurements for accuracy, from eight participants performing five tasks on two treatments (undocumented and annotated), and 16 ($8 * 2$) measurements for time, from two treatments.
	We compare accuracy and time measurements between treatments using the non-parametric two-sided Wilcoxon Signed Rank (WSR) test as the measurements between treatments are paired and the sample size is small.
	In addition, all measurements are analyzed based on descriptive statistics.
	Figure \ref{fig:combined_graph} shows the box-plot of accuracy scores, time measurements, and perception ratings.

	\textbf{Time.}
	Both mean and median values of time taken for the annotated treatment (\textit{mean=336.6s, median=328.5s}) are lower than the undocumented variants (\textit{mean=486.4s, median=464.5s}).
	Moreover, WSR test on time measurements showed statistical significance (\textit{p-value=0.025, statistic=34.0}).
	The large difference in completion time for the undocumented variant is associated with the back-and-forth navigation in the notebook trying to find relevant areas.
	This shows that participants took significantly more time to complete comprehension tasks when given an undocumented notebook.
	
	\textbf{Accuracy.}
	The mean accuracy of all comprehension tasks was greater for the annotated treatment, except for task T6, where it was equal.
	The variance of accuracy across the tasks was three times higher for the undocumented treatment, showing that it is more likely to yield better accuracy with annotated notebooks.
	However, the median is greater for the annotated treatment only in T4 and T5.
	In addition, WSR test showed that the accuracy scores from the study are not statistically significant between two treatments (\textit{p-value=0.106, statistic=55.0}).
	Nonetheless, note that the study was not time-boxed. Participants thus took significantly longer to solve the tasks correctly for undocumented notebooks.
	
	\textbf{Navigability and Usefulness.}
	The perceived ratings for statements in Table \ref{table:likertstatements} showed that the participants find \toolname considerably helpful in completing the tasks.
	None of the participants disagreed to statements S1, S2 and S4, and none of them agreed to statement S3.
	All participants showed interest in actually installing the tool when it is published.
	
	\textbf{Qualitative Results.}
	Participants noted that \toolname would be especially useful when dealing with very large undocumented notebooks as it provides a ``map'' of the notebook.
	Participants also found the function documentation to be useful, given that the libraries are continuously evolving and that they would often come across methods that they have not seen before.
	Furthermore, minor recommendations to improve the taxonomy categories were noted and added to the final version.
	Recommendations to change the layout of the plugin were also noted and will be considered in future versions.

	\textbf{Threats to Validity.}
	The study we conducted is prone to some common limitations of conducting user studies.
	Due to the small number of participants, it may not be representative of a larger population.
	However, participants were selected from all fields: students, professionals, and academics to get inputs from different perspectives.
	Furthermore, since the study follows a within-subject design, the order of tasks and treatments can have an effect on the outcome.
	Therefore, to limit the learning effect, we use latin-square design to randomize the order of treatments, tasks, and multiple choices.
	However, using notebooks that only use the Keras API might have had a learning effect as the study progressed.
	Although the participants were experienced working with the default \jnb environment, \toolname adds additional interfaces that might seem confusing at first.
	As a result, some participants did not make full use of \toolname's capabilities.
	
	\section{RQ2: Accuracy of Callsite Recognition}
	\label{sec:RQ2}
	\textbf{Micro-benchmark Results.} 
	We evaluate \toolname{} for complete and sound recognition of callsites.
	The analysis is complete when there are no false positives, and sound when there are no false negatives.
	In total, the analysis is sound in 113 of 121 cases, and complete in 113 of 121 test cases.
	Lack of soundness in eight of 121 test cases are due to the lack of implementation for analyzing challenging \python{} features such as decorators.
	On the other hand, out of the eight test cases that are incomplete, only three of them are due to missing implementation of challenging features.
	The remaining five test cases are not complete because our analysis is context-insensitive. As a result, it over-approximates the solution in certain scenarios.

	Note that we do not perform a direct comparison of \toolname with PyCG because the micro-benchmark does not pose specific challenges to flow-sensitivity, except for the new \textit{flow\_sensitive} category with eight test cases that we added.
	Furthermore, note that PyCG does not output line numbers in its analysis and therefore a direct automated comparison is not possible.
	When manually compared to PyCG for the \textit{flow\_sensitive} category, as expected, PyCG is incomplete for all eight test cases.
		
	\textbf{Real-world Benchmark Results.} 
	Table \ref{table:realworldbench_results} lists the precision and recall values of \toolname{} for real-world notebooks.

	\toolname{} achieves an average of 95.6\% precision and 95.3\% recall.
	Note that in four instances, the analysis achieves 100\% precision and recall.

	The precision loss is due to our type-stub database's over-approximation of return-types.
	For instance, a call \texttt{x.isnull()} can be either \texttt{Series.isnull} or \texttt{DataFrame.isnull}, depending on whether  \texttt{x} is a \textit{Series} or \textit{Dataframe}, which is determined based on the underlying structure of the data.
	However, this is not straightforward to infer and needs advanced data-flow analysis.

	Where recall is lost, it is because our analysis lacks supports for some complex \python{} features.
	
\begin{table}[]
	\renewcommand{\arraystretch}{1.2}
	\centering
	\caption{Real-world benchmark evaluation of \toolname for callsite recognition and header annotation}	
	\label{table:realworldbench_results}
	\begin{tabular}{crrrr}
		\toprule
		\multirow{2}{*}{\textbf{Id}} & \multicolumn{2}{c}{\textbf{Callsite Recognition}} & \multicolumn{2}{c}{\textbf{Header Annotation}} \\
		\cmidrule(lr){2-3} \cmidrule(lr){4-5}
		& {\textbf{Precision}} & {\textbf{Recall}} & {\textbf{Precision}} & {\textbf{Recall}} \\ 
		\midrule
		1                            & 100                           & 90.0                                                     & 71.4                                                        & 100                                                      \\
		2                            & 100                           & 100                                                      & 84.2                                                        & 100                                                      \\
		3                            & 95.8                          & 95.8                                                     & 100                                                         & 87.5                                                     \\
		4                            & 100                           & 100                                                      & 75.8                                                        & 95.9                                                     \\
		5                            & 93.8                          & 97.4                                                     & 83.3                                                        & 90.9                                                     \\ \midrule
		6                            & 86.0                          & 97.4                                                     & 100                                                         & 100                                                      \\
		7                            & 94.4                          & 91.9                                                     & 83.3                                                        & 88.2                                                     \\
		8                            & 100                           & 100                                                      & 94.1                                                        & 76.2                                                     \\
		9                            & 90.9                          & 97.2                                                     & 83.6                                                        & 92.7                                                     \\
		10                           & 100                           & 100                                                      & 85.0                                                        & 90.7                                                     \\ \midrule
		11                           & 97.1                          & 85.0                                                     & 68.8                                                        & 100                                                      \\
		12                           & 96.0                          & 96.0                                                     & 94.7                                                        & 94.7                                                     \\
		13                           & 94.3                          & 94.4                                                     & 100                                                         & 100                                                      \\
		14                           & 94.4                          & 93.5                                                     & 74.1                                                        & 87.0                                                     \\
		15                           & 91.3                          & 91.3                                                     & 87.5                                                        & 87.5                                                     \\ 
		\midrule
		\textbf{Average}             & \textbf{95.6}                          & \textbf{95.3}                                                     & \textbf{85.7}                                                        & \textbf{92.8}                                                     \\ 
		\bottomrule
	\end{tabular}
\end{table}

\textbf{Extended Real-world Benchmark Results.} 
Table \ref{table:extendedbench_results} lists the precision and recall values of \toolname{} for real-world notebooks.
\toolname{} achieves an average of 94.4\% precision and 91.6\% recall.
In two instances, the analysis achieves 100\% precision and recall.
We observe similar results compared to our evaluation with the real-world benchmark.

However, in the instance of ``nb\_111547", a notably low recall rate of 66.7\% is observed.
This issue arises from the authors' frequent use of the \texttt{DataFrame.apply} and \texttt{Series.apply} methods within the notebook.
These methods accept a function reference as an argument and apply it across the DataFrame's contents. 
While \toolname can partially analyze this functionality, the behavior of these functions is dependent on the data type within the DataFrame, information that is not statically available.
This highlights a limitation of our approach.

\begin{table}[]
	\renewcommand{\arraystretch}{1.2}
	\centering
	\caption{Extended benchmark evaluation of \toolname for callsite recognition and header annotation}	
	\label{table:extendedbench_results}
	\begin{tabular}{rrrrr}
		\toprule
		\multicolumn{1}{c}{\multirow{2}{*}{\textbf{Id}}} & \multicolumn{2}{c}{\textbf{Callsite Recognition}} & \multicolumn{2}{c}{\textbf{Header Annotation}} \\
		\cmidrule(lr){2-3} \cmidrule(lr){4-5}
		& {\textbf{Precision}} & {\textbf{Recall}} & {\textbf{Precision}} & {\textbf{Recall}} \\ 
		\midrule
\textbf{nb\_111547}                                       & 95.4                      & 66.7                  & 82.5                    & 91.7                 \\
nb\_12952                                        & 98.1                      & 98.7                  & 73.4                    & 100                  \\
nb\_13973                                        & 100                       & 100                   & 92.0                    & 76.7                 \\
nb\_2004                                         & 88.9                      & 94.1                  & 95.8                    & 74.2                 \\
nb\_24325                                        & 97.2                      & 100                   & 95.5                    & 91.3                 \\ \midrule
nb\_27915                                        & 92.0                        & 95.8                  & 74.5                    & 100                  \\
nb\_44070                                        & 85.7                      & 88.9                  & 81.8                    & 93.1                 \\
nb\_50450                                        & 89.2                      & 87.9                  & 79.0                    & 89.1                 \\
nb\_5433                                         & 97.7                      & 80.8                  & 92.5                    & 100                  \\
nb\_6005                                         & 84.4                      & 85.7                  & 56.4                    & 91.7                 \\ \midrule
nb\_62235                                        & 95.8                      & 90.7                  & 93.9                    & 91.2                 \\
nb\_62337                                        & 100                       & 100                   & 100                     & 97.5                 \\
nb\_79900                                        & 97.2                      & 89.7                  & 92.9                    & 96.3                 \\
nb\_87231                                        & 99.1                      & 99.1                  & 85.7                    & 81.1                 \\
nb\_96079                                        & 95.6                      & 95.6                  & 76.3                    & 93.5                 \\ \midrule
\multicolumn{1}{c}{\textbf{Average}}             & \textbf{94.4}             & \textbf{91.6}         & \textbf{84.8}           & \textbf{91.2}        \\ \bottomrule
	\end{tabular}
\end{table}

\section{RQ3: Accuracy of Generated Headers}
\label{sec:RQ3}
\toolname uses identified function calls in code cells to automatically add relevant headers based on the taxonomy of ML operations.
We evaluated the headers generated by \toolname{} for precision and recall against manually annotated headers.
Again, we use our real-world benchmark as a basis.

\textbf{Results.}
The resulting precision and recall are listed on the right side of Table \ref{table:realworldbench_results} and Table \ref{table:extendedbench_results}.
The headers that are generated by \toolname are matched on the high-level categories of the taxonomy listed in Figure \ref{t:ml_operations}.
\toolname{} achieves a precision of 85.7\% and recall of 92.8\% on the real-world benchmark and 84.8\% precision and 91.2\% recall on the extended real-world benchmark.
\toolname{} achieves similar results with the extended real-world benchmark.
Precision is lost because some functions can be mapped to more than one ML operation. 

\textbf{Impact of Pattern Matching.}
To study the impact of pattern matching on the accuracy of header annotations, we employed \toolname with pattern matching deactivated and compiled the resultant data.
Observations indicate that, within the context of a real-world benchmark, activating pattern matching enhances the recall of header annotation by 10.6\%, and for the extended real-world benchmark, recall improves by 14.8\%.
Conversely, the precision of header annotation diminishes across both benchmarks due to the over-approximation of patterns that are mapped to multiple ML operations.
Specifically, for the real-world benchmark, precision decreased by 2.3\%, and for the extended real-world benchmark, it decreased by 4.6\%.

\section{RQ4: Dependency of HeaderGen on Type Information}
\label{sec:eval_typeevalpy}

HeaderGen relies on its analyzer’s ability to infer variable types to accurately resolve callsites.
To systematically measure the influence of type inference on the accuracy of HeaderGen, we conduct a comprehensive comparison of type inference tools within the \typeevalpy framework. 
We compare \toolname against \emph{PyCG}, \emph{pyright}, \emph{Jedi}, and \emph{HiTyper} using our real-world benchmark.
Note that our type-stub database of ML libraries was supplied to \emph{pyright} and \emph{Jedi} to facilitate their analysis.

We employ \typeevalpy to generate statistics related to exact type matches using its micro-benchmark.
Table \ref{tab:typeevalpy_comparisons} displays the exact matches for all the tools being compared. 
Additionally, we evaluated \emph{HiTyper}, which is a hybrid analysis technique using deep learning for type inference.
This method leverages \emph{PyCG} as a static analysis tool within its process.

\begin{table}
	\renewcommand{\arraystretch}{1.5}
	\centering
	\scriptsize
	\setlength\tabcolsep{3.25pt}
	\caption{Exact matches of type inference tools for micro-benchmark categories. \textit{FR:} Function return types, \textit{FP:} Function parameter types, and \textit{LV:} Local variable types, \textit{845:} Total type annotations, \textit{154:} Total test cases.}
	\label{tab:typeevalpy_comparisons}
	\begin{tabular}{lrrr|rrr|rrr|rrr|rrr}
		\toprule
		\multicolumn{1}{c}{\multirow{2}{*}{\textbf{Category}}} & \multicolumn{3}{c|}{\textbf{Total}}                                       & \multicolumn{3}{c|}{\textbf{HeaderGen}}                                   & \multicolumn{3}{c|}{\textbf{Jedi}}                                        & \multicolumn{3}{c|}{\textbf{Pyright}}                                     & \multicolumn{3}{c}{\textbf{HiTyper}}                                     \\ 		\cmidrule(lr){2-4} \cmidrule(lr){5-7} \cmidrule(lr){8-10} \cmidrule(lr){11-13} \cmidrule(lr){14-16}
		\multicolumn{1}{c}{}                                   & \multicolumn{1}{c}{FR} & \multicolumn{1}{c}{FP} & \multicolumn{1}{c|}{LV} & \multicolumn{1}{c}{FR} & \multicolumn{1}{c}{FP} & \multicolumn{1}{c|}{LV} & \multicolumn{1}{c}{FR} & \multicolumn{1}{c}{FP} & \multicolumn{1}{c|}{LV} & \multicolumn{1}{c}{FR} & \multicolumn{1}{c}{FP} & \multicolumn{1}{c|}{LV} & \multicolumn{1}{c}{FR} & \multicolumn{1}{c}{FP} & \multicolumn{1}{c}{LV} \\ \midrule
		args                                                   & 17                     & 10                     & 16                     & 17                     & 9                      & 12                     & 12                     & 0                      & 9                      & 8                      & 1                      & 8                      & 12                     & 0                      & 6                      \\
		assignments                                            & 22                     & 4                      & 56                     & 15                     & 1                      & 33                     & 20                     & 0                      & 21                     & 20                     & 0                      & 25                     & 21                     & 4                      & 9                      \\
		\rowcolor{black!5}\textbf{builtins}                                               & 3                      & 4                      & 61                     & 0                      & 0                      & 26                     & 0                      & 0                      & 21                     & 1                      & 0                      & 45                     & 1                      & 2                      & 18                     \\
		classes                                                & 41                     & 8                      & 76                     & 39                     & 7                      & 67                     & 0                      & 0                      & 57                     & 1                      & 0                      & 46                     & 27                     & 2                      & 41                     \\
		decorators                                             & 29                     & 17                     & 12                     & 11                     & 6                      & 2                      & 10                     & 0                      & 8                      & 7                      & 0                      & 3                      & 8                      & 0                      & 3                      \\
		dicts                                                  & 23                     & 3                      & 82                     & 23                     & 3                      & 60                     & 21                     & 0                      & 34                     & 19                     & 2                      & 50                     & 20                     & 3                      & 22                     \\
		direct\_calls                                          & 10                     & 3                      & 11                     & 10                     & 3                      & 8                      & 6                      & 0                      & 7                      & 3                      & 0                      & 6                      & 3                      & 2                      & 4                      \\
		\rowcolor{black!5}\textbf{dynamic}                                                & 1                      & 0                      & 8                      & 1                      & 0                      & 2                      & 1                      & 0                      & 2                      & 1                      & 0                      & 2                      & 1                      & 0                      & 5                      \\
		exceptions                                             & 0                      & 0                      & 2                      & 0                      & 0                      & 2                      & 0                      & 0                      & 1                      & 0                      & 0                      & 1                      & 0                      & 0                      & 1                      \\
		external                                               & 3                      & 1                      & 12                     & 0                      & 0                      & 3                      & 0                      & 0                      & 8                      & 0                      & 0                      & 2                      & 1                      & 0                      & 4                      \\
		functions                                              & 11                     & 9                      & 17                     & 8                      & 9                      & 12                     & 5                      & 0                      & 14                     & 5                      & 2                      & 13                     & 5                      & 5                      & 5                      \\
		generators                                             & 13                     & 8                      & 35                     & 9                      & 4                      & 17                     & 5                      & 0                      & 23                     & 4                      & 3                      & 18                     & 10                     & 5                      & 11                     \\
		imports                                                & 3                      & 0                      & 22                     & 3                      & 0                      & 11                     & 1                      & 0                      & 16                     & 3                      & 0                      & 20                     & 3                      & 0                      & 11                     \\
		kwargs                                                 & 8                      & 6                      & 8                      & 8                      & 5                      & 5                      & 7                      & 0                      & 4                      & 4                      & 0                      & 5                      & 7                      & 0                      & 0                      \\
		lambdas                                                & 7                      & 7                      & 20                     & 3                      & 7                      & 4                      & 6                      & 0                      & 11                     & 2                      & 0                      & 1                      & 3                      & 0                      & 7                      \\
		lists                                                  & 17                     & 4                      & 38                     & 14                     & 1                      & 26                     & 17                     & 0                      & 27                     & 13                     & 0                      & 25                     & 16                     & 3                      & 16                     \\
		mro                                                    & 16                     & 0                      & 18                     & 14                     & 0                      & 16                     & 0                      & 0                      & 13                     & 0                      & 0                      & 14                     & 15                     & 0                      & 11                     \\
		returns                                                & 15                     & 4                      & 24                     & 11                     & 1                      & 16                     & 11                     & 0                      & 17                     & 9                      & 0                      & 13                     & 10                     & 1                      & 5                      \\
		\midrule
		\multicolumn{1}{c}{\multirow{2}{*}{\textbf{Totals}}}   & \textbf{239}           & \textbf{88}            & \textbf{518}           & \textbf{186}           & \textbf{56}            & \textbf{322}           & \textbf{122}           & \textbf{0}             & \textbf{293}           & \textbf{100}           & \textbf{8}             & \textbf{297}           & \textbf{163}           & \textbf{27}            & \textbf{179}           \\  \cmidrule(lr){2-4} \cmidrule(lr){5-7} \cmidrule(lr){8-10} \cmidrule(lr){11-13} \cmidrule(lr){14-16}  
		\multicolumn{1}{c}{}                                   & \multicolumn{3}{c}{\textbf{845}}                                         & \multicolumn{3}{c}{\textbf{564}}                                            & \multicolumn{3}{c}{\textbf{415}}                                         & \multicolumn{3}{c}{\textbf{405}}                                         & \multicolumn{3}{c}{\textbf{369}}                                        
		
		\\ \bottomrule
	\end{tabular}
\end{table}

\textbf{Results.}
\toolname exhibits dominant performance across various categories. 
Specifically, it showcases superior type inference for function returns in categories such as \textit{args}, \textit{classes}, \textit{decorators}, \textit{dicts}, \textit{direct\_calls}, \textit{imports}, \textit{kwargs}, \textit{lists}, and \textit{mro}.
Notably, in the \textit{classes} category, \toolname outperformed other tools in function returns.
\toolname also shows robust performance in inferring local variables and function parameters across multiple categories, highlighting its comprehensiveness.

Other tools have promising results in certain categories:
\emph{Jedi} has its strengths in inferring function returns for \textit{decorators} and excels in \textit{lambdas} with better inference of function returns and local variables.
\emph{pyright} shows prowess in the builtins category for local variables.
\emph{HiTyper} exhibits strength in categories like \textit{assignments} for function returns and \textit{generators} where it excels in recognizing function returns and local variables.

While \toolname performed robustly in most categories, some outliers exist.
In the \textit{builtins} category, \toolname failed to infer function returns, making it a notable exception.
Similarly, in \textit{dynamic} and \textit{external} categories, its performance wasn't comprehensive.

In addition, it is notable that both \emph{Jedi} and \emph{pyright} consistently fail to infer function parameter types across all categories.
This design choice stems from their tendency to label the type annotation of function parameters as ``\texttt{Any}'', reflecting Python's duck typing philosophy.
Essentially, this means a non-type annotated function can accept a variable of any type. 
While this might align with Python's flexibility, it is not beneficial for static analysis of Python code.

\textbf{Discussion.}
The standout performance of \emph{Jedi} and \emph{pyright} in the \textit{builtins} and \textit{external} categories provide insights into their underlying design.
Academic tools, such as \emph{PyCG}, overlook the use of user provided type hints, called typestubs, when analyzing the types of Python elements.
Conversely, both \emph{Jedi} and \emph{pyright} have fine-tuned their processes to incorporate these typestubs effectively.
Nevertheless, the design paradigms of both \emph{Jedi} and \emph{pyright} are oriented towards specific use cases, such as auto-completion and Integrated Development Environments (IDEs) integration, influencing the nature of their outputs.
Accessing their internal static analysis data, like points-to and type information, is not straightforward for a more generalized source code analysis, as needed for \toolname.

To achieve better performance, \toolname combines \emph{PyCG's} algorithm, EAG, and leverages typestubs, where required, to increase accuracy.
However, the support for handling typestubs are more mature in \emph{Jedi} and \emph{pyright} and this is evident by their performance in the highlighted categories where typestubs are required the most.

Another observation is the performance of \emph{HiTyper}.
Despite its foundation on \emph{PyCG} and its hybrid analysis approach, its varied performance on different categories are difficult to explain considering its reliance on ML techniques.
We posit that incorporating \toolname's analysis might enhance the accuracy of \emph{HiTyper}. 
We aim to explore this in future research.

\textbf{Modeling of Pandas Behavior.}
Listing~\ref{lst:typingschallenges} shows simplified data manipulation methods of the Pandas library based on our real-world benchmark.
Furthermore, Table \ref{table:tools_type_inference} lists the type of each variable used in Listing~\ref{lst:typingschallenges} as inferred by the tools being compared.
It can be seen that \emph{pyright}, \emph{Jedi}, and \emph{HiTyper} fail to infer return-types of variables \texttt{x1} through \texttt{x6}.
This is because \toolname can model complex pandas accesses while the other tools fail.
For instance, in line 6, a dot notation access \texttt{df.a} is ignored by other tools while \toolname models it as a \texttt{Series}.

\begin{lstlisting}[language=python, numbers=right,label=lst:typingschallenges,caption=Common uses of Pandas DataFrame that existing tools fail to infer.,style=pstyle,float=h]
	import pandas as pd
	
	df = pd.read_csv("./input.csv")
	x1 = df["a"].map(lambda x: x + 1.0)
	x2 = df.iloc[[False]].reset_index().copy()
	x3 = df.a.fillna(0)
	x4 = df.groupby(["a"])[["b"]].agg({"b": ["min"]})
	x5 = df[["b", "c"]]
	x6 = df.c.values.astype(int)
\end{lstlisting}

\begin{table}[h]
	\renewcommand{\arraystretch}{1.2}
	\centering
	\caption{Comparison of type inference by existing tools for listing \ref{lst:typingschallenges}}	
	\label{table:tools_type_inference}	
	
	\begin{tabularx}{\textwidth}{cXXXXX}
		\toprule
		\textbf{Variable} & \textbf{Actual} & \textbf{HeaderGen} & \textbf{Pyright} & \textbf{Jedi} & \textbf{HiTyper} \\ 
		\midrule
		df  & DataFrame & DataFrame & DataFrame & DataFrame & List \\ 
		x1  & Series    & Series    & Any       & Any       & Int       \\ 
		x2  & DataFrame & DataFrame & Any       & Any       & Int       \\ 
		x3  & Series    & Series    & Any       & Any       & Int       \\ 
		x4  & DataFrame & DataFrame & Any       & Any       & Int       \\ 
		x5  & DataFrame & DataFrame & Any       & Any       & Int       \\ 
		x6  & Ndarray   & Ndarray   & Any       & Any       & Int       \\ 
		\bottomrule
	\end{tabularx}
\end{table}

\section{RQ5: Comparison with Existing Tools}
\label{subsec:RQ5}
We conduct a comparison of \toolname against \emph{PyCG}, \emph{pyright}, \emph{Jedi}, and \emph{HiTyper} focusing on callsite recognition and header annotation, using our real-world benchmark.
Since \emph{pyright} and \emph{Jedi} are primarily configured for type checking and auto-completion, we incorporated auxiliary functions to gather type information and callsite data for comparison. 

\begin{table}[]
	\renewcommand{\arraystretch}{1.2}
\centering
\caption{Comparative evaluation of existing tools on real-world benchmark for callsite recognition and header annotation}
\label{table:tools_call_sites_compare}    
	\begin{tabular}{lSSSS}
		\toprule
		\multicolumn{1}{c}{\multirow{2}{*}{\textbf{Tool}}} & \multicolumn{2}{c}{\textbf{Callsite Recognition}}                            & \multicolumn{2}{c}{\textbf{Header Annotation}}                               \\
		\cmidrule(lr){2-3} \cmidrule(lr){4-5}
& \textbf{Precision} & \textbf{Recall} & \textbf{Precision} & \textbf{Recall} \\
\midrule
		\multicolumn{5}{c}{\textbf{Real-world Benchmark}}                                                                                                                                                                \\ \hline
		HeaderGen                                          & 95.6                                   & 95.3                                & 85.7                                   & 92.8                                \\
		Pyright                                            & 96.7                                   & 87.2                                & 83.8                                   & 82.7                                \\
		Jedi                                               & 84.6                                   & 65.8                                & 85.1                                   & 69.8                                \\
		PyCG                                               & 41.7                                   & 23.3                                & 84.6                                   & 26.2                                \\ \midrule
		\multicolumn{5}{c}{\textbf{Extended Real-world Benchmark}}                                                                                                                                                       \\ \midrule
		HeaderGen                                          & 94.4                                   & 91.6                                & 84.8                                   & 91.2                                \\
		Pyright                                            & 95.7                                  & 82.4                                & 87.4                                   & 73.1                                \\
		Jedi                                               & 86.6                                   & 70.9                                & 89.2                                   & 58.8                                \\
		PyCG                                               & 62.0                                   & 43.4                                & 80.3                                   & 7.6                                
		 \\ \bottomrule 
	\end{tabular}
\end{table}

\textbf{Results.}
The precision and recall values for both the real-world benchmarks and extended real-world benchmarks are listed in Table~\ref{table:tools_call_sites_compare}.
Since header annotation is based on identified callsites, it is evident that higher recall of callsite recognition leads to higher recall in header annotation.
\toolname achieves the highest callsite recall of $95.3\%$ which leads to a $92.8\%$ recall in header annotation of code cells.
However, \emph{pyright} is the closest with $87.2\%$ recall for callsite recognition which leads to $82.7\%$ recall for header annotation.
Note that without our type-stub database, these tools would perform even worse.
We observe a similar trend on the extended real-world benchmark results with \toolname leading with a callsite recall of $91.6\%$ and $91.2\%$ recall in header annotation of code cells. The loss of precision is attributed to the over-approximation of return-types in our type-stub database as discussed earlier.

Note that incorporating HiTyper into \toolname for an automated evaluation of callsite recognition, based on HiTyper's inferred types, proved challenging.
The difficulty originated from the process of translating HiTyper's output back to the original source code and further integrating it with \toolname's expected format.
Despite significant effort, we opted to discontinue this integration process.
Nonetheless, insights from \typeevalpy, as outlined in Section~\ref{sec:eval_typeevalpy}, suggests that HiTyper's performance may not surpass that of \toolname.
This perspective was further reinforced by our manual review of HiTyper's results when applied to the real-world benchmark notebooks.

\section{Related Work}
\label{sec:relatedwork}

\textbf{Tool-support for Jupyter Notebooks.}
Over the past few years, numerous studies \citep{keryStoryNotebookExploratory2018, ruleExplorationExplanationComputational2018, pimentelLargeScaleStudyQuality2019, koenzenCodeDuplicationReuse2020, wangBetterCodeBetter2020, eppersonStrategiesReuseSharing2022a, quarantaBestPractices2022, grotovLargescaleComparisonPython2022} have examined coding patterns in notebooks. 
These studies consistently indicate a deficiency in the quality of notebooks, signaling a need for greater attention from the software engineering community.
Despite these findings, there is a noticeable gap in research efforts focusing on tools that can address the identified issues.

In a step towards addressing this gap, \cite{wangDocumentationMattersHumanCentered2022a} introduced \textit{Themisto}.
This tool prompts data scientists to document their code cells. 
It employs a deep learning method to auto-generate code documentation in natural language and then suggests to users whether to integrate or directly apply this documentation. 
Notably, \emph{Themisto} relies on the Abstract Syntax Tree (AST) of \python code for model training, without incorporating SA methods to extract more contextual information from the code.
We posit that the analytical capabilities of \toolname might enhance deep learning approaches, potentially yielding better outcomes.

In another study, \cite{pimentelLargeScaleStudyQuality2019} studied 1.4 million notebooks for features that affect reproducibility and suggested a set of best practices.
Following this, \cite{wangAssessingRestoringReproducibility2020} propose \textit{Osiris}, a tool-based approach to restore reproducibility in \jnb{}s by using AST parsing for data-flow analysis to find dependencies of variables between code cells.
Furthermore, \cite{yang2022data} design a SA approach to detect data leakage in notebooks. 
In contrast, our contribution to this domain focuses on the automatic annotation of code cells, offering tool-support for literal programming.

\textbf{Static Analysis for Python.}
Python has grown to be one of the leading programming languages, but the field still observes a significant lack in SA tools, as highlighted by a study on Python's features by~\cite{yangComplexPythonFeatures2022}. 
Yang et al. emphasize that Python's distinct characteristics make it difficult to use traditional analysis methods developed from existing scientific research.
One reason for this is Python's dynamic features, like duck typing, which while beneficial for rapid prototyping, complicate its analysis.

Only as recently as 2021 was a notable breakthrough achieved in the domain of SA for Python: the introduction of \emph{PyCG}, a method for constructing call graphs~\citep{salisPyCGPracticalCall2021a}.
Yet, this method does not account for flow of values and is not tailored for Jupyter notebooks. Furthermore, Python still lacks a comprehensive SA framework to produce data flow intermediate representations.
The most related work in this area is the Scalpel project \citep{liScalpelPythonStatic}, but it too falls short, particularly in inferring return types for external functions and considering notebook cells.

In addition, \cite{monat2020static} delve into using static abstract interpretation to aid Python type analysis.
This approach covers a broad range of constructs and precisely combines domains, allowing sound knowledge of nominal and structural types and exceptions raised in the program.
Building on this foundation, \cite{monat2020value} developed \emph{MOPSA}, a prototype tool that integrates value analysis.
However, MOPSA focuses specifically on analyzing Python code that are combined with C code.
In our study, we aim to enhance the current landscape of SA for Python in practical scenarios.
We present methods for resolving return types of external library APIs and extracting flow-sensitive function call sites, leveraging def-use relations.
However, MOPSA has the potential to provide more reliable type stubs from C modules, which can benefit our work and other analyses.
We suggest future research to study this more closely.

\textbf{Code classification.}
Code classification is fundamental for various tasks, such as determining code authorship, identifying the programming language, and understanding source file content.
While significant work exists in this domain, studies that specifically focus on Jupyter notebooks are limited.
Several machine learning techniques have been instrumental in the progress of this field.
Methods like Max-entropy~\citep{zevin2017machine}, decision tree~\citep{ugurel2002s}, K-nearest neighbor (KNN), and Naive Bayes~\citep{barstad2014predicting} are noteworthy.
They are proficient in tasks like predicting the programming language of source code and identifying the topics within a document.
Additionally, recent advancements in deep learning have introduced weakly supervised transformer-based architectures for the classification and tagging of source code, as highlighted by \cite{zhang2022coral}.

In a recent contribution to this domain, \cite{ramasamyWorkflowAnalysisData2022} presented a supervised method specifically tailored for data science code classification.
They approached the task as a topic classification challenge and explored both single-label (one label per cell) and multi-label classification to account for multiple data science operations within a single code cell.
However, there are inherent limitations with their supervised classification approach.
As noted by the authors, the classification of notebook cells—particularly those associated with evaluation, prediction, and visualization tasks—can achieve lower F1-scores due to the skewed distribution of their training dataset, where certain labels are underrepresented.
In contrast, our method bypasses the need for an extensive pre-training process. 
By grounding our approach in API usage analysis, we aim to provide a solution that is more resilient and applicable in real-world contexts.

\section{Limitations \& Future work}
\label{sec:discussion}

To enhance the accuracy of function name resolution, we leverage Python's reflection mechanism.
This, however, may limit the coverage of APIs, contingent on the version of the library installed.
To counteract this, we are considering the use of static API mapping techniques to address transitive imports in Python.
Our current analysis predominantly focuses on machine learning applications.
Nonetheless, the architecture of the framework we designed is not confined to this domain.
\toolname is equipped to annotate notebooks, accommodating domain-specific return-type stubs and library classifications.
Lastly, the present version of \toolname utilizes an ML taxonomy that comprises three main categories.
Recent studies on notebooks have introduced a more detailed taxonomy, and future iterations of \toolname will adopt this refined classification.

The taxonomy classification model has been developed using a small dataset consisting of 400 function calls, which represents a constraint of our methodology.
The process of curating a dataset is costly and requires the involvement of several experts. 
In an attempt to enhance generalization, we are presently investigating methodologies for employing our dataset to fine-tune a large language model for this classification task.

Additionally, input from \toolname can be used to automatically restructure code cells in \jnb{}s for better readability.
It can achieve this by reorganizing complex code cells, particularly those encompassing multiple ML operations, into a sequential arrangement.
Furthermore, the efficient and accurate function call analysis provided by \toolname paves the way for large-scale mining studies of \python code base.

\section{Conclusion}
\label{sec:conclusion}

In numerous instances, \jnb{}s found in practical settings lack documentation, leading to challenges in understanding and navigating them.
Addressing this, \toolname employs accurate static analysis to automatically annotate \jnb{}s with structural headers.
These headers are derived from a categorization of machine learning operations.
Notably, \toolname displayed high precision and recall when tested against both micro and real-world benchmarks.
Our evaluations against expert-curated ground truth further confirmed \toolname's capability to annotate headers with adequate precision and high recall.
Furthermore, when assessing the type inference capabilities of various static analysis tools using the \typeevalpy framework, \toolname consistently surpassed its counterparts. 
To understand \toolname's real-world impact, we conducted a user study.
The results indicated that ML practitioners perceive \toolname as a valuable tool for enhancing both program understanding and navigation.

\section*{Data Availability Statements}

Data to reproduce experiments described in Sections \ref{sec:userstudy}, \ref{sec:RQ2}, \ref{sec:RQ3},\ref{sec:eval_typeevalpy}, and \ref{subsec:RQ5} along with the source code for \toolname is published on GitHub at:\\ \url{https://github.com/secure-software-engineering/HeaderGen}.\\

Data to reproduce Table \ref{tab:typeevalpy_comparisons} along with source code for \typeevalpy is published on GitHub at:\\ \url{https://github.com/secure-software-engineering/TypeEvalPy}

\section*{Conflict of interest}

 The authors declare that they have no conflict of interest.

\bibliographystyle{spbasic}      
\bibliography{bibfile.bib,bibfile2.bib}

\end{document}